# Exploring the Origins of Anti-Ambipolarity in BBL Polymer: Links to Redox Chemistry, Electronic Structure, and Structural Dynamics

**Maryam Ghotbi,[1] Alejandro Aviles,[2] and Perla B. Balbuena[1,2,3,*]**

**[1]Department of Chemistry, Texas A&M University, College Station, TX, US**

**[2]Department of Chemical Engineering, Texas A&M University, College Station, TX, US**

**[3]Department of Materials Science and Engineering, Texas A&M University, College Station, TX, USA**

*** e-mail: balbuena@tamu.edu**

## Abstract

We examine the intrinsic physical-chemical properties of the conjugated ladder-type polymer poly(benzimidazobenzophenanthroline) (BBL) in response to electron transfer. We aim at explaining the origin of the antiambipolar behavior behind the observed BBL nonlinear response associated with specific device architectures. To elucidate this point, we use theory and computation based on first principles, including density functional theory optimizations, ab initio molecular dynamics, time-dependent DFT, and Marcus-theory analysis.  Our results reveal that this redox response is not simply monotonic, but follows an alternating odd/even pattern in which gap narrowing and reopening occur sequentially before near-gapless behavior emerges at high charging. Converging theoretical evidence in this work demonstrates that bell shaped conductivity in BBL originates in its fundamental electronic structure and supramolecular organization.

## 1-Introduction

The rapid expansion of artificial intelligence and data-intensive computing has created a pressing need for hardware that can process information with the efficiency and adaptability of biological neural systems. [1,2] This need extends to emerging applications at the interface of electronics and biology, ranging from brain-inspired computing and neural signal processing to biometric authentication based on individual brainwave signatures.[3-5] In all of these domains, conventional von Neumann architectures helping with separation of memory from computation, incur prohibitive energy costs associated with data movement and cannot replicate the in-material signal integration that biological systems perform naturally. [2,6,7]

Neuromorphic hardware requires materials whose conductance evolves non-linearly and non-monotonically with electrochemical bias, mimicking the activation-inactivation dynamics of biological ion channels. [6, 8, 9] Device-level strategies to achieve such behavior have often relied on anti-ambipolar architectures, where a non-monotonic current–voltage response emerges from engineered heterojunctions and circuit design.[10, 11] These approaches have enabled multivalued logic, artificial synapses, and energy-efficient neuromorphic circuits, positioning anti-ambipolarity as a key functional motif in emerging electronics. [11] However, in such systems, the non-linear response is typically imposed by device architecture rather than arising intrinsically from the material itself. [12-14].

Organic mixed ionic-electronic conductors provide a complementary paradigm in which ionic and electronic degrees of freedom are intrinsically coupled through electrochemical doping. [15, 16]. In these systems, bulk ion penetration directly modulates the electronic structure, enabling reversible and dynamic control over conductivity. [17] Despite this coupling, many conjugated polymers exhibit monotonic increases in conductivity with doping, indicating that non-linear transport behavior is not generally encoded at the molecular level. [15] This raises a fundamental question: **can anti-ambipolarity emerge intrinsically from molecular electronic structure and redox chemistry, rather than device architecture alone?**

The ladder-type polymer poly(benzimidazobenzophenanthroline) (BBL) provides a compelling system in which this distinction becomes critical. In organic electrochemical transistors (OECTs), BBL exhibits a stable, reversible antiambipolar transfer characteristic, where conductivity first increases and then decreases with electrochemical doping. [18] Earlier spectroelectrochemical studies of BBL revealed multiple doping-induced conductivity maxima and alternating insulating and conducting states, indicating that non-monotonic transport behavior is intrinsically linked to its redox chemistry.[19] This Gaussian response has been exploited to realize conductance-based organic electrochemical neurons capable of spiking at biologically relevant frequencies. [18] Importantly, unlike conventional anti-ambipolar devices based on heterojunctions, this behavior arises within a single-phase material under electrochemical control, suggesting that the origin of the non-monotonic conductance is rooted in intrinsic electronic structure and redox chemistry. [12, 13]

At the microscopic level, this behavior is governed by the evolution of charge carriers and electronic states. Polarons formed at low doping support efficient transport, whereas multiply charged states formed at higher doping reduce carrier mobility. [20, 21] Spectroscopy provides a powerful, non-destructive probe of electronic structure through the interaction of matter with electromagnetic radiation, enabling direct characterization of charge-transfer states and excitation processes, particularly in heterogeneous systems. [22]. In situ spectroscopic and electrical investigations reveal the sequential formation of polaronic in-gap states, manifested as characteristic near- and mid-infrared absorption features and the emergence of occupied states near the Fermi level. [21] Complementary theoretical and spectroscopic studies further demonstrate that these states correspond to distinct redox species, including polarons and bipolarons, which govern the alternating conductive and insulating regimes in BBL. [23] These measurements, supported by high conductivities (~2 S $cm^{-1}$) and low activation energies (~44 meV), indicate delocalized charge transport along the planar ladder backbone. [21] Complementary theoretical studies further show that multi-electron states adopt unconventional configurations, including triplet polaron pairs, reflecting a complex interplay between charge localization, spin, and electronic structure. [24] More broadly, redox-active molecular systems can support charge disproportionation and interfragment charge transfer, generating multiple accessible electronic states that can mediate information transfer via switching. [25-28] Despite these insights, a quantitative connection between redox-dependent electronic structure, excitation character, and non-monotonic transport remains unresolved.

Here, we demonstrate that the antiambipolar response of BBL is intrinsically encoded in its chemistry. Early theoretical work showed that charge transfer in BBL modifies both the backbone geometry and the electronic gap, with doping driving the naphthalenic unit toward a more quinoid-like structure and reducing the band gap.[29] In contrast, our results reveal that this redox response is not simply monotonic, but follows an alternating odd/even pattern in which gap narrowing and reopening occur sequentially before near-gapless behavior emerges at high charging.[30] Bader charge analysis identifies carbonyl groups as the primary redox-active sites, establishing the molecular origin of charge accumulation. Time-dependent DFT reveals a transition from delocalized inter-monomer charge-transfer excitations at intermediate doping to localized excitations at high charge density, consistent with the spectroscopically observed evolution of electronic states. Constrained-DFT Marcus calculations show that electron transfer rates peak at intermediate charge states before collapsing in the inverted regime, providing a kinetic origin for the conductance maximum. Ab initio molecular dynamics further demonstrates that structural reorganization enhances inter-chain coupling at optimal doping, while higher charge densities induce disorder that suppresses transport.

Together, these results establish that anti-ambipolarity in BBL is not solely a device-level phenomenon but arises directly from its redox chemistry, electronic structure, and structural dynamics.

## 2. Computational Methods

### 2.1. Periodic Density Functional Theory: Electronic Structure and Geometry Optimization

Periodic DFT calculations were performed using the Vienna Ab Initio Simulation Package (VASP, version [6.3.2]). [31] The projector-augmented wave (PAW) method was employed to describe core-valence interactions. [32] A plane-wave kinetic energy cutoff of 520 eV and convergence criterion of $10^{-8}$ eV were used throughout. Geometry optimizations were performed with ISIF = 2, allowing atomic positions to relax under fixed cell shape and volume and spin polarization was included in all calculations. The Brillouin zone was sampled using a Γ-centered 3 × 2 × 6 k-point mesh. Methfessel–Paxton smearing with σ = 0.05 eV was applied. Dispersion interactions were accounted for using the DFT-D3 many-body dispersion correction for structural relaxation runs. [33]

To identify the appropriate exchange-correlation functional for subsequent electronic structure analysis, the bandgap of trans-BBL was benchmarked against the experimentally reported value of ~1.9 eV using three levels of PBE-based theory:[34] the semilocal PBE functional, [35] the dispersion-corrected PBE-MBD approach [36, 37], and the PBE0 hybrid functional. [38] PBE0 calculations employed the Normal algorithm with IMIX = 1 and AMIX = 0.4. Only PBE0 reproduced the experimental bandgap with high fidelity; it was therefore adopted for all subsequent electronic-structure analyses (see Supporting Information, Section S1, for the full functional comparison).

### 2.2. Density of States

Total and projected density of states (TDOS/PDOS) were computed from single-point SCF calculations using the PBE0 hybrid functional on previously relaxed geometries. [38] Hybrid DOS calculations employed the Damped algorithm. Orbital-resolved projections were obtained and fragment-resolved PDOS was analyzed using VASPKIT package to resolve contributions from various structural groups across redox states. [39]

### 2.3. Implicit Solvation

Environmental screening effects were assessed by repeating selected electronic structure calculations within an implicit aqueous dielectric continuum using the VASPsol solvation model with a relative permittivity of ε = 78.4, representative of bulk water at room temperature.[40] All other settings were identical to the corresponding gas-phase calculations, enabling a direct comparison of gas-phase and solvated electronic structures.

### 2.4. Bader Charge Analysis

Bader charge partitioning was performed on converged PBE0 charge densities using the grid-based algorithm of Henkelman and coworkers. [41] Bader charges were evaluated for the pristine BBL unit cell and for cells with one, two, and three added electrons, with a uniform compensating background charge applied to maintain periodicity. Atom-resolved charge differences relative to the neutral system were used to construct charge redistribution heatmaps, identifying the primary redox-active sites along the backbone.

### 2.5. Ab Initio Molecular Dynamics

Ab initio molecular dynamics (AIMD) trajectories were generated using VASP in the NVT ensemble with a Nosé-Hoover thermostat at T = 300 K. [42] A Γ-only k-point sampling (1 × 1 × 1) was used to reduce computational cost for the extended supercell. Electrons were added sequentially in increments of 0.5 e/RU from 0 to 3.0 e/RU, with a compensating uniform background charge applied at each step. Simulations were performed for pristine BBL as well as systems containing explicit $K^+$ counterions and neutral KCl ion pairs. Structural analysis including bond-length statistics, RMSD values, and chain-orientation metrics was carried out by averaging over trajectory windows after discarding an initial equilibration period of [0.1] ps.

### 2.6. Molecular Geometry Optimizations and Spin-State Analysis

Gas-phase geometry optimizations of the BBL trimer across multiple charge and spin states were performed using Gaussian 16. [43] The range-separated hybrid functional ωB97X-D [44] was employed with the split-valence basis set 6-311++G(2d,2p) applied to all standard atoms. Optimizations were performed for charge states spanning +1 h/RU to 2 e/RU in increments of one electron, covering singlet through [quintet/specified] spin multiplicities to identify the lowest-energy spin configuration at each redox level.

### 2.7. Time-Dependent DFT - Electronic Excitations

Vertical excitation energies and transition character were computed using linear-response TDDFT as implemented in Gaussian 16, using a 50/50 exchange mixing scheme with the ωB97X-D functional and the 6-311++G(2d,2p) basis set. [45] All TDDFT calculations were performed on the gas-phase BBL trimer at geometries optimized for the corresponding charge and spin state. A total of 35-300 excited states were computed per redox configuration. Electron and hole density distributions for each excitation were analyzed using the Multiwfn program [46, 47] to generate atom-resolved hole/electron density maps and quantify inter-fragment charge transfer (IFCT) character across the three BBL monomers (M1, M2, M3).

### 2.8. Marcus Electron Transfer Rates - Constrained DFT

Electron transfer parameters were computed using the constrained DFT (cDFT) approach as implemented in NWChem. [48] Calculations employed the range-separated hybrid functional ωB97X-D3 with the 6-311G(d,p) basis set. Open-shell systems were treated with spin-unrestricted DFT with tight integral thresholds enforced. Level shifting was disabled to ensure stable convergence of the constrained electronic states. The direct Coulomb algorithm was used throughout.

For each symmetry-distinct monomer-to-monomer hop within the BBL trimer across reduction states n = 1-6, two constrained electronic states were defined by confining excess charge to the donor and acceptor monomers respectively, thereby imposing a localized single-electron transfer scenario. Geometry optimizations were performed for each constrained state to obtain the relaxed donor (reactant) and acceptor (product) structures. The thermodynamic driving force (ΔG) and reorganization energy (λ) were extracted from the four total energies obtained at the donor and acceptor geometries in both electronic states, following the standard Marcus four-point scheme. Electronic coupling matrix elements (H) were computed from separate cDFT electron transfer calculations using the cDFT-CI approach, [49] in which the coupling is evaluated directly from the overlap and Hamiltonian matrix elements between the two diabatic cDFT states. Semiclassical Marcus rate constants were evaluated at T = 298.15 K using: [50, 51]

$$k_{et} = (2\pi/\hbar) \cdot H^2 \cdot (4\pi\lambda k_BT)^{-1/2} \cdot \exp[-(\Delta G + \lambda)^2 / (4\lambda k_{BT})]$$

All 2.8 rate constants are reported in $s^{-1}$. Hops yielding $k_{et} < 10^2\ s^{-1}$ are classified as kinetically frozen and attributed to deep Marcus inverted-regime conditions ($|\Delta G| \gg \lambda$).

## 3. Discussion

### 3.1. Bader Charge Analysis: Redox-Site Identification and Charge Redistribution

Fragment-resolved Bader charge analysis reveals that electron uptake in **poly(benzimidazobenzophenanthroline) (BBL)** is strongly nonuniform and evolves systematically with successive reduction. Using a chemically meaningful partitioning into the **naphthalenic segment**, **carbonyl region**, and **benzimidazole moiety**, the first added electron is distributed mainly over the benzimidazole moiety and the naphthalenic fragment, with total charge changes of about −0.404 and −0.361 e, respectively, while the carbonyl region already accommodates a substantial fraction of the added charge (−0.236 e). Upon further reduction, however, the fragment contributions diverge: the naphthalenic part localizes progressively less negative charge (−0.305 at e2 and −0.260 at e3), whereas the carbonyl region becomes increasingly electron rich (−0.283 at e2 and −0.287 at e3). Because this fragment contains only four atoms, it also shows the largest charge accumulation on a per-atom basis, increasing from about −0.059 e $atom^{-1}$ at e1 to approximately −0.071 to −0.072 e $atom^{-1}$ at e2–e3. This identifies the carbonyl-containing unit as the most intense **local redox-active motif** in the repeat unit. At the same time, the benzimidazole moiety remains the largest overall charge reservoir and becomes slightly more negative with increasing electron count (−0.404, −0.412, and −0.453 e for e1–e3), indicating that reduction is not confined to the carbonyls but extends over the adjacent N-containing conjugated framework.

At the atomistic level, the carbonyl carbons **C19** and especially **C20** exhibit the clearest and most persistent negative shifts, whereas the associated oxygens **O31** and **O32** remain negative but do not increase as strongly. This suggests that the excess charge is accommodated not on oxygen atoms alone, but over the broader **carbonyl π-manifold**, with significant participation of the adjacent conjugated backbone. Overall, the data support a reduction mechanism in which the first electron is more broadly distributed over the aromatic framework, while subsequent electrons drive a redistribution toward a more polarized state characterized by enhanced carbonyl-centered localization superimposed on continued delocalization over the benzimidazole-containing segment. This chemically selective redistribution is consistent with a later evidence that modification of the naphthalenic unit in BBL measurably alters its optical and electronic properties, underscoring the sensitivity of charge accommodation to local backbone structure.[52] In this picture, the naphthalenic fragment participates most strongly at low reduction but becomes progressively less involved as the charge density increases.

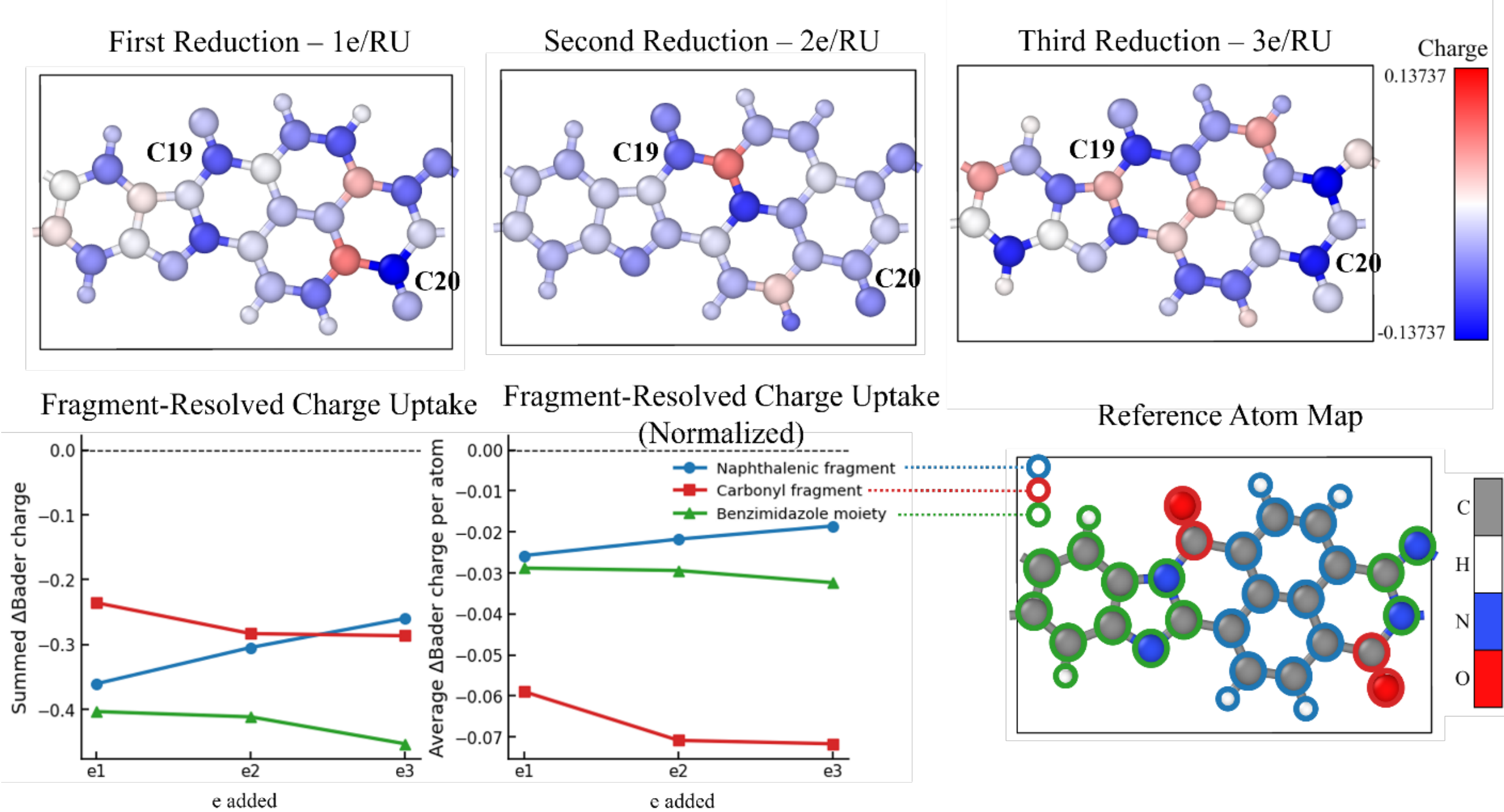

***Figure 1.*** *Evolution of Bader charge redistribution in trans-BBL upon sequential electron addition.* ***Top row:*** *atom-resolved heat maps for the changes in atomic charge at 1e, 2e, and 3e reduced states.* ***Bottom row:*** *fragment-resolved charge changes relative to the neutral state,* $\Delta q = q_n - q_{n-1}$*, shown as total fragment charge (left) and average charge per atom (right). The data reveal strongly site-selective charge uptake, with increasing localization in the carbonyl-containing region and benzimidazole moiety as reduction proceeds.*

Although BBL is experimentally characterized as an n-type material,[53] its intrinsic electronic structure responds to both reduction and oxidation in a strongly non-monotonic manner. [29] To evaluate this behavior independent of device-specific transport factors, we analyze the redox dependence of the total density of states (TDOS) together with the bandgap extracted from the calculated electronic structure.

Previous experimental and DFT studies of BBL thin films reported a neutral-state band gap in the range of ~1.7–1.9 eV, [29, 54]consistent with the narrow-gap semiconducting character of the polymer, as is also illustrated by the calculated neutral bandgap at 0 e/RU in Figure 2. Upon reduction, the TDOS shows a clear redistribution of spectral weight toward the Fermi level, accompanied by a zigzag evolution of the bandgap with electron count (Figure 2). At 1 electron per repeating unit (1 e/RU), the bandgap becomes very small, consistent with the emergence of states close to the Fermi level. Increasing the reduction to 2 e/RU leads to a reopening of the bandgap to ~0.6 eV, indicating a reorganization of the occupied and unoccupied manifolds rather than a monotonic closing with increasing electron count. Upon further reduction to 3 e/ru, the bandgap again becomes very small, and the TDOS exhibits substantial spectral weight at the Fermi level, consistent with an effectively gapless electronic structure within the resolution of the calculation which further in in this document is shown to be electronically unstable.

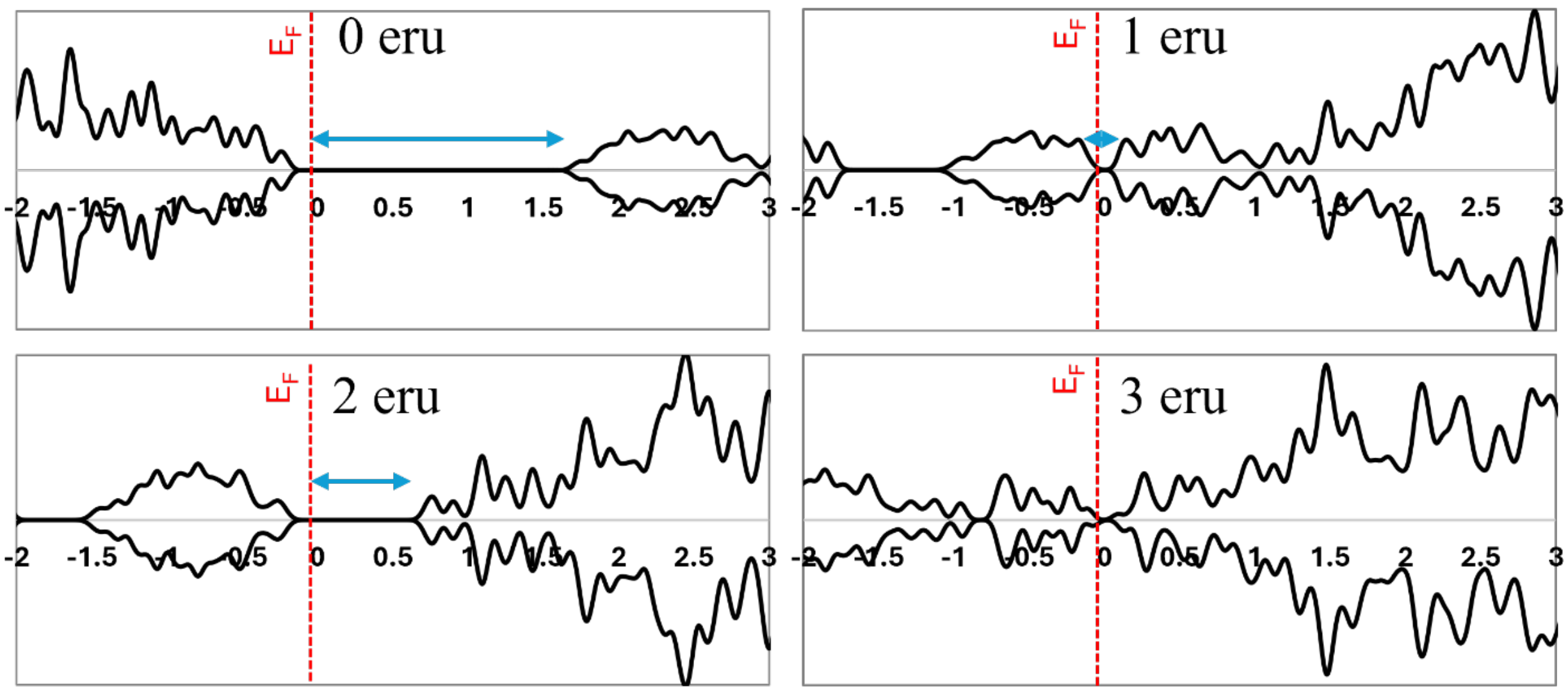

***Figure 2.*** *TDOS of neutral (0 eru = 0 "electrons per repetitive unit"), singly (1 eru), doubly (2 eru), and triply (3 eru) reduced trans-BBL, showing the relationship between redox state and bandgap. The neutral and doubly reduced states retain relatively large gaps, while the singly and triply reduced states exhibit strongly narrowed bandgaps.*

A closely analogous non-monotonic response is observed upon oxidation. Illustrated in Figure 3, The neutral system exhibits a bandgap of ~1.6 eV, whereas oxidation to 1 hole per repeating unit (1 $h^+$/RU) collapses the gap to ~0.1 eV, indicating a dramatic shift of electronic states toward the Fermi level. Oxidation to 2 $h^+$/RU increases the bandgap again to ~0.8 eV, demonstrating that the second oxidation step stabilizes a distinct electronic configuration with greater energetic separation between occupied and unoccupied states. At 3 $h^+$/RU, the bandgap nearly vanishes and the TDOS approaches a continuous distribution across the Fermi level, indicating a transition to a near-gapless electronic structure.

The redox level of trans-BBL induces a non-monotonic bandgap evolution. Odd and even charge states (both for electron addition and removal) are characterized by alternating gap narrowing and reopening prior to gap collapse at higher charge densities. Importantly, the presence of the same zigzag trend under both reduction and oxidation shows that this behavior is an intrinsic electronic-structure response to successive charging, rather than a polarity-specific feature that would be expected solely from n-type transport considerations.

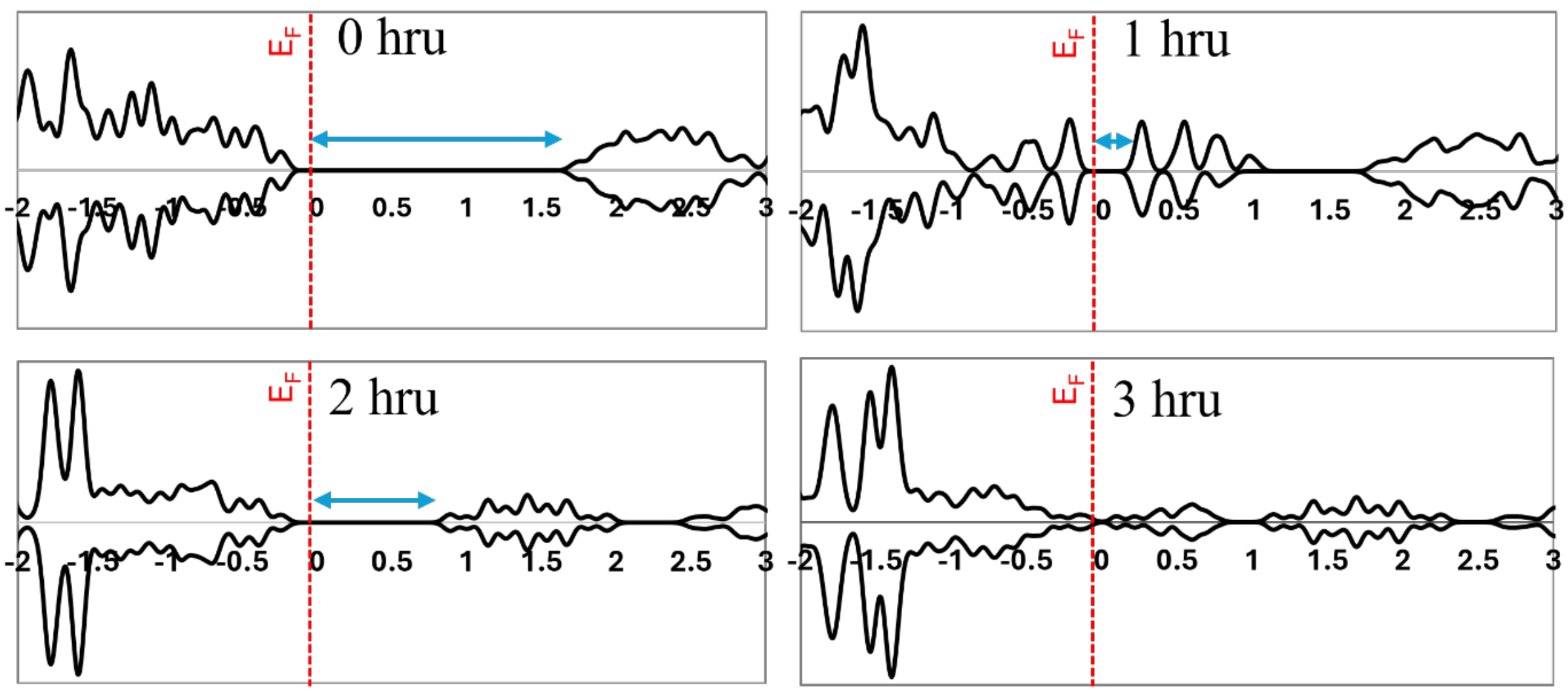

***Figure 3.*** *TDOS of BBL in the neutral state (0 hru = 0 "holes per repetitive unit") and upon oxidation to 1, 2, and 3 $h^+$/RU. Oxidation produces a nonmonotonic electronic response, with strong bandgap narrowing at 1 $h^+$/RU, partial reopening at 2 $h^+$/RU, and an almost continuous distribution across the Fermi level at 3 $h^+$/RU.*

### 3.2. Spin-State Stability and Its Implications for Electron Transfer in BBL

The rate of electron transfer is another key materials property that can directly influence conductivity. As a first step toward evaluating electron transfer behavior, we identified the most stable electronic structures of BBL by comparing the total energies of a BBL trimer across multiple spin multiplicities and redox states. Calculations were performed for charge states ranging from +1 hole per repeating unit ($h^+$/RU) to 2 electrons per repeating unit ($e^-$/RU), in increments of one electron.

Across most redox states, the low-spin configuration was found to be the most stable. The only exceptions occur for the trimer with 2 and 4 added electrons, for which a triplet spin multiplicity is energetically favored. Notably, BBL, with its experimentally observed n-type character, [24] exhibits greater thermodynamic stability in its reduced states compared to oxidized ones, with the lowest total energy obtained for the triplet state at 0.66 $e^-$/RU.

The neutral system displays the largest HOMO-LUMO gap, consistent with the maximum bandgap observed in the crystalline phase. However, this correspondence does not persist across the other redox states, indicating that molecular-level HOMO-LUMO gaps do not map one-to-one onto crystal bandgaps as the charge state changes. In part, this reflects the different levels of description: the molecular calculations treat a finite BBL trimer, while the bandgaps are derived from periodic calculations that explicitly include intermolecular interactions and band formation in the crystal.

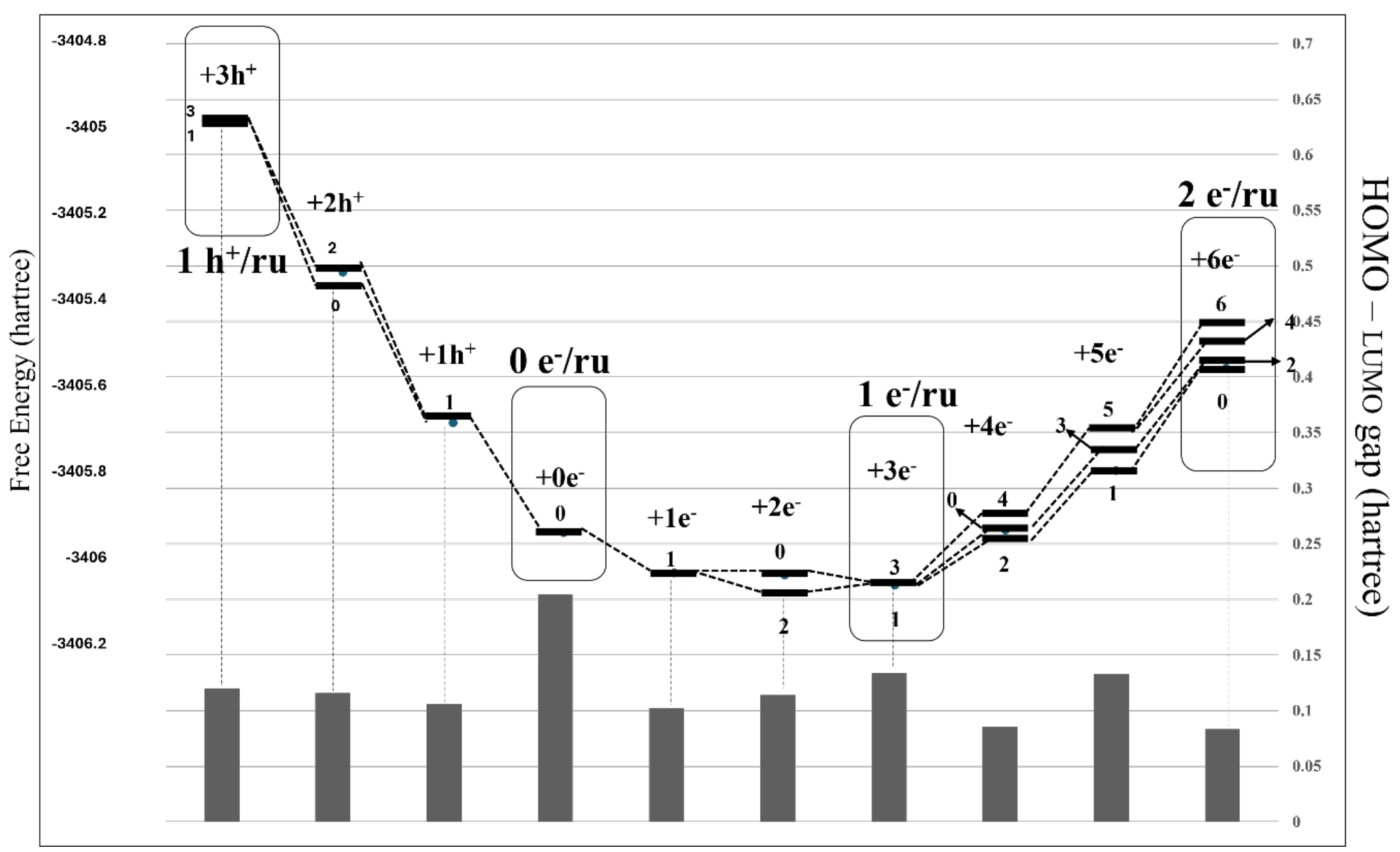

***Figure 4.*** *Relative stability of the BBL trimer across charge states ranging from +1 h/ru to 2 e/ru. Consistent with its n-type character, BBL preferentially stabilizes electron-localized states, whereas hole-containing states are energetically much less favorable.*

### 3.3. Fragment Resolved DOS Reveals Cooperative Charge Delocalization in BBL

Analysis of the density of states (DOS) provides a powerful framework for tracing the electronic origin of the redox-dependent transport behavior observed in BBL. The fragment-resolved partial DOS (PDOS) reveals that the frontier electronic structure is backbone-dominated across all reduction levels, with the carbonyl groups playing an increasingly significant but secondary role upon successive electron addition. In the neutral state, the occupied frontier region is governed primarily by the benzimidazole moiety, with a lesser contribution from the naphthalenic fragment; the C=O groups are conspicuously absent from the highest occupied bands, while the unoccupied manifold already carries noticeable carbonyl character (Figure 5). Reduction to 1 and 2 e/RU substantially enhances carbonyl participation near the frontier, signaling the progressive activation of a carbonyl-centered acceptor character embedded within an otherwise backbone-based electronic network; consistent with the large binding energy shifts observed at the carbonyl carbon in XPS measurements of alkali-metal-doped BBL films.[21] At 3 e/RU, the low-energy unoccupied states intensify and shift downward, accompanied by a renewed dominance of backbone character, as a result of a more spatially delocalized reduced electronic structure. Comparison of the reduced and neutral states shows that electrochemical reduction in BBL is not confined to a single functional group. Instead, the added charge is shared between the rigid conjugated backbone and the carbonyl acceptor sites, producing a distributed polaronic state across the monomer. This behavior differs from simple site-localized charging models and is consistent with the extended polaron delocalization suggested by the low activation energies measured in doped BBL films. [21]

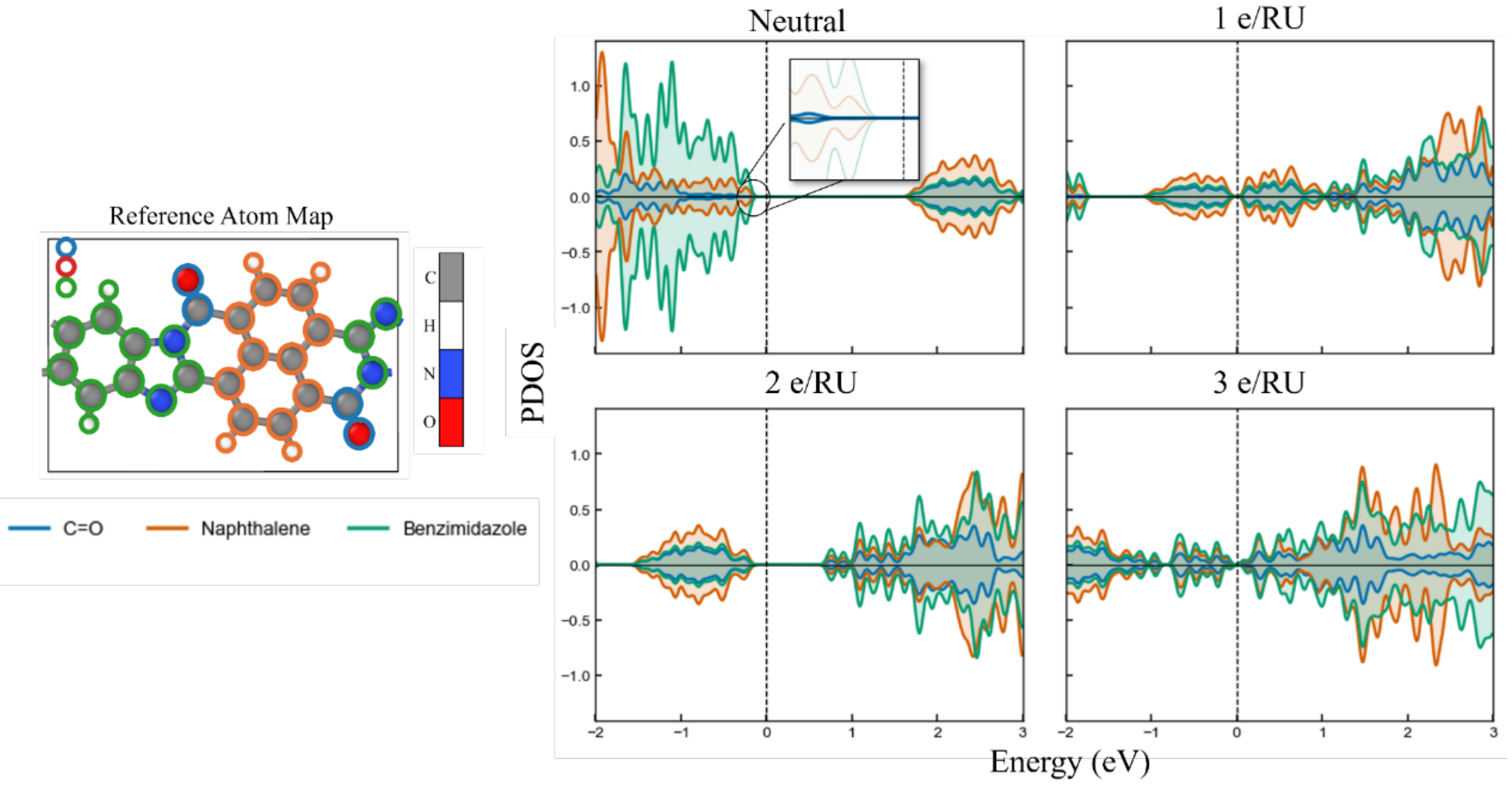

***Figure 5.*** *Partial Density of States for neutral, singly, doubly, and triply reduced BBL. C=O activation emerges at the first reduction step. Carbonyl character is present in all near-Fermi DOS regions, except in the occupied manifold of the neutral state. Contributions of fragments tend to be similar at reduced states, regardless of number of atoms in each.*

### 3.4. Redox-Dependent Excitation Dynamics

Time-dependent DFT calculations on the BBL trimer reveal that the character of the lowest electronic excitation undergoes a systematic, redox-driven reorganization. The trend illustrates a delocalized excitonic state in the neutral system, through a series of asymmetric inter-monomer charge-transfer configurations at low and intermediate reduction, to a fully localized single-unit excitation at high electron loading. These excitation topology transitions provide a molecular basis for the nonlinear and memory-like conductance responses that underpin the neuromorphic and antiambipolar functionality of BBL.

In the neutral trimer, the first excitation is distributed across all three repeat units with dominant weight on the central monomer (M2) as shown by bright regions in Figure 6, bottom row. Fragment-resolved analysis (Figure 6, top row) reveals an intrinsic donor-acceptor partitioning within each repeat unit: carbonyl groups contribute preferentially to the electron distribution while nitrogen atoms contribute to the hole, establishing an internal polarity that primes the system for directional charge redistribution upon excitation.

Upon reduction, the excitation acquires pronounced inter-monomer charge-transfer (CT) character that evolves continuously and non-monotonically with carrier density. At 0.33 e/RU, a well-defined M2 → M3 CT transition emerges, with the hole concentrated on M2 and the electron localized on M3 (Fig. S3). At 0.66 e/RU, this polarization shifts to an M1/M2-centered arrangement, with the electron predominantly on M2 and the hole distributed across M1 and M2 (Fig. 6-top). At 1.00 e/RU, M2/M3 coupling reasserts itself (Fig. S3), while by 1.33 e/RU the dominant redistribution pathway reverses direction, with excitation weight shifting further toward M3 (Fig. 6-top). Throughout this low-to-intermediate reduction regime, the backbone remains the primary carrier of excitation character while carbonyl participation grows steadily with added charge. This is consistent with a cooperative backbone–carbonyl response in which charge redistribution remains electronically flexible and directional. The continuous reorientation of CT pathways with carrier density means that modest electron injection does not localize the excitation but instead generates multiple competing inter-unit redistribution channels, providing a molecular origin for the analog tunability and nonmonotonic current response central to neuromorphic operation.

A qualitative crossover occurs at 1.66 e/RU (Fig. S3), where both hole and electron densities collapse predominantly onto a single repeat unit (M1), signaling the breakdown of extended inter-monomer redistribution. This transition is not a simple intensification of the lower-doping behavior; it represents a reorganization of both the spatial topology and the fragment identity of the excitation. At 2.00 e/RU, nitrogen character increases sharply on the electron side, and by 3.00 e/RU the excitation is almost entirely confined to M1 and becomes strongly nitrogen-dominated, with carbonyl participation reduced to a secondary role. Figures S3 and S4 provide a comprehensive view of the donor and acceptor contributions at the atomic level (Figure S3) and fragment level (Figure S4).

The upward/downward trend of transition state delocalization establishes redox state as an internal control parameter governing both the spatial extent and the fragment character of the first excitation in BBL. Low and intermediate reduction sustain asymmetric, CT-active states distributed across multiple repeat units, supporting the electronically flexible charge redistribution and competing transport channels required for antiambipolar behavior. High reduction drives a transition to a fragment-confined, nitrogen-centered excitation topology that promotes charge retention and self-limiting transport, consistent with the stabilization of localized polaronic states implicated in non-volatile memory function. This behavior shows a direct relation to experimentally observed bell-shaped conductivity [18], suggesting chemical explanations for antiambipolarity in BBL.

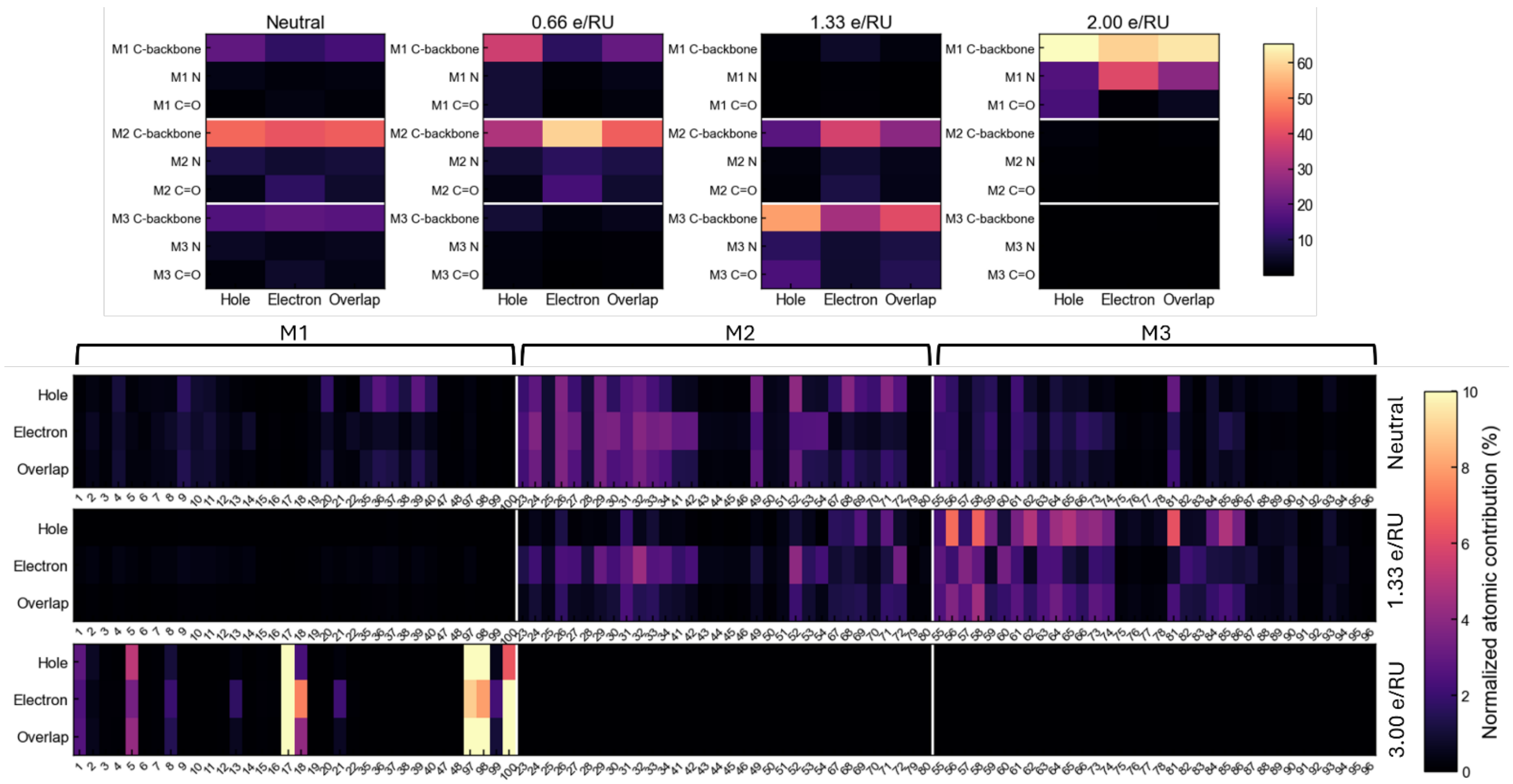

***Figure 6.*** *Distribution of Hole and Electron in various redox states of BBL trimer, obtained with TD-DFT.* ***Top row:*** *Fragment-resolved hole and electron distribution for carbon backbone, N atoms and carbonyl groups of each monomer.* ***Bottom row:*** *Atomic contributions of hole and electron in the first excited state for then neutral, mildly doped, and highly doped BBL, showing low, high and extremely low delocalization, in order. Tilted values on the X-axis show atomic indices.*

### 3.5. Marcus Electron Transfer Rates and Optimal Doping

Recent device-level studies of BBL heterojunctions show that charge transport in the BBL film is governed by thermally activated hopping. [53] In this context, our Marcus-theory analysis provides a microscopic description of the hopping process within BBL itself, linking the redox-state dependence of electronic coupling, reorganization energy, and driving force to the observed transport response, that is in perfect agreement with previous electronic excitations results for various doping levels of BBL.

The established excitation topology provides a natural interpretive framework for the computed electron transfer kinetics. The CT-active, multi-pathway regime identified by TDDFT at low and intermediate reduction (0.33-1.33 e/RU) corresponds to conditions under which Marcus theory predicts the fastest monomer-to-monomer hopping rates; conversely, the onset of excitation localization at 1.66 e/RU anticipates a kinetic regime in which long-range charge redistribution becomes increasingly suppressed. To quantify this correspondence, we computed electron transfer rate constants ($k_{et}$) for all symmetry-distinct single-electron hops within the BBL trimer across reduction states n = 1-6, using constrained DFT to extract the driving force (ΔG), reorganization energy (λ), and electronic coupling (H) for each transition.

While intramolecular charge transfer provides an initial framework for the intrinsic behavior of BBL, Marcus theory further elucidates the relationship between charge state and conductive response. Given that conductivity $\sigma = ne\mu$, an analysis of $k_{et}$ is essential for defining mobility μ. Although H exhibits high sensitivity to local geometry and electron distribution, the observed kinetic trends are remarkably consistent with previously reported experimental benchmarks. In general, the more symmetric electronic distributions tend to be more favorable, as illustrated by thicker arrows in Figure 7, left. For the mono-reduced trimer (n = 1), the single-electron transfer rate is on the order of $10^8$ $s^{-1}$. The introduction of the second and third electrons significantly enhances $k_{et}$ by 5 to 6 orders of magnitude, effectively transitioning the BBL trimer into a high-mobility state which is consistent with the electronically flexible, CT-active excitation character identified in this doping window. The suppression of $k_{et}$ observed for selected n = 4 pathways is attributable to the Marcus inverted regime, (Figure S7-left) wherein |ΔG| up to ~6 eV far exceeds λ ~ 1 eV, rendering most hopping channels kinetically inaccessible despite the increased carrier density. The n = 5 state restores kinetic facility to ~$10^{13}$ $s^{-1}$, suggesting optimal electron transport at approximately 1.3 e/RU, in alignment with performance metrics observed in device-level BBL. Ultimately, the sixth electron (2 e/RU) saturates the system, collapsing $k_{et}$ to ~$10^6$ $s^{-1}$. This value lies lower than the singly reduced state and in direct correspondence with the nitrogen-centered, spatially confined excitation topology identified by TDDFT at this doping limit. Figure S-7 summons the relationships of Marcus parameters for all calculated states. The convergence of two independent theoretical frameworks at the same critical threshold of ~2 e/RU provides strong mutual support that excitation localization and kinetic saturation arise from the same underlying electronic reorganization. This connection offers an atomistic explanation for the antiambipolar transport behavior of BBL.

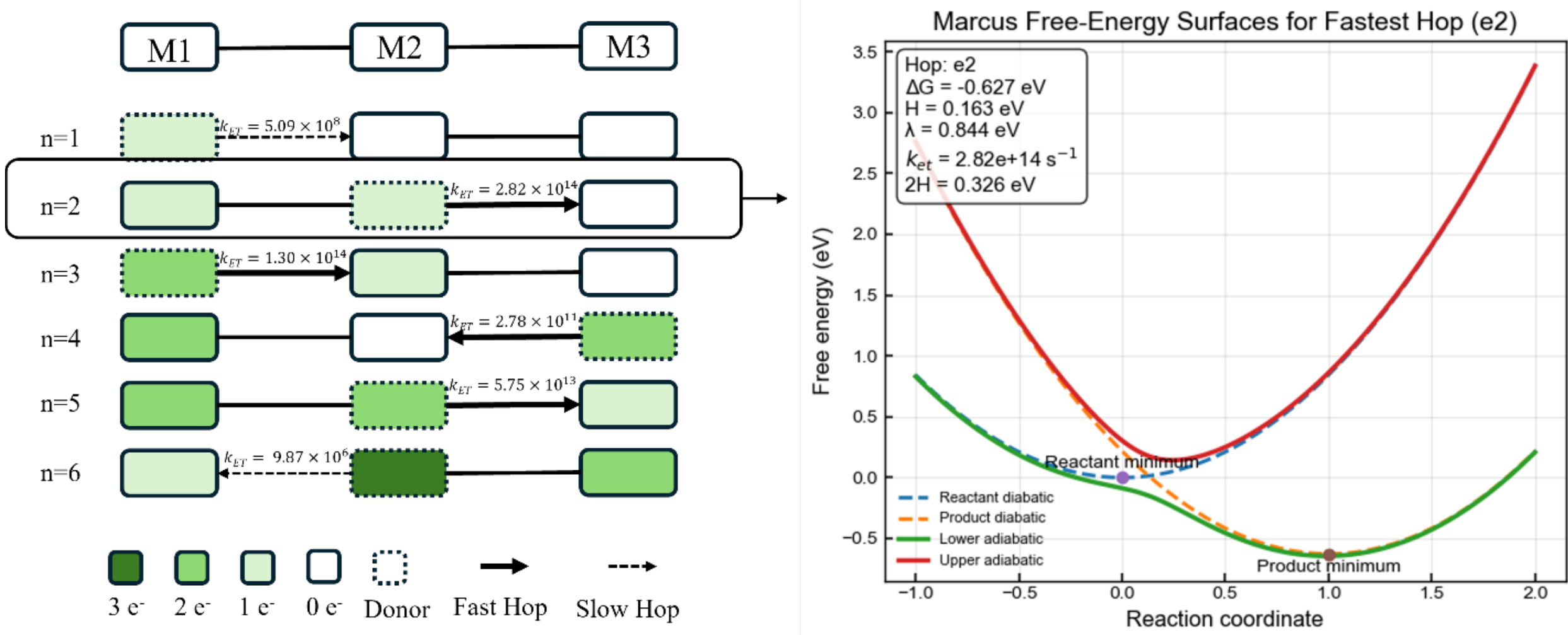

***Figure 7.*** *Growth of electron occupancy with reduction and the eventual collapse of efficient hopping channels at high doping, consistent* ***Left:*** *Dominant single-electron hopping pathways in the BBL trimer (M1–M2–M3) for reduction states* $n = 1$ *to 6. Green boxes denote occupied monomers, arrows indicate the fastest hop in each state, and annotated* $k_{et}$ *values show the corresponding transfer rates; solid arrows mark normal-regime hops and dashed orange arrows mark suppressed or inverted-regime hops.* ***Right:*** *Marcus free-energy diagram for the fastest computed hopping pathway, showing reactant and product diabatic parabolas (dashed) and the corresponding adiabatic surfaces including electronic coupling* $H$ *(solid).*

### 3.6. Structural responses to doping contribute to antiambipolarity

The structural response of the BBL chains to progressive electrochemical reduction reveals a compelling and non-trivial reorganization pathway. In the ab-initio Molecular Dynamics (AIMD) trajectory of the pristine cell, addition of the first 4 electrons (corresponding to 0.5 e/RU) leaves the chain registry largely intact: the stacks maintain their crystalline order with only minimal tilting and a slight lateral offset relative to the underlying layer (Figure 8-left). This suggests that at low-to-moderate doping levels, the rigid ladder backbone effectively resists the structural perturbation introduced by the added charge.

Upon the subsequent two electron additions, however, the singly reduced chains undergo a more pronounced lateral displacement, progressively shifting until the inter-chain distance matches that of the bottom layer. Inspection of the periodically replicated supercell reveals an emergent long-range order: the chains reorganize into pairs exhibiting near-perfect π-orbital overlap (Figure 8, 1e/RU). This paired, high-overlap stacking motif is not a disordering of the crystal but rather a reorganization into a new structural phase.

The electronic consequences of this structural reorganization are substantial and directly quantifiable. The high-overlap stacking geometry that emerges near 1.5 e/RU is in strong agreement with prior computational results indicating that inter-chain charge transport is most effective in this doping regime. To quantify the effect of stacking geometry on charge transport, we calculated the electronic coupling, H, for a single electron hop along the polymer stacking direction using two representative trajectory snapshots. When the lateral chain offset decreases by 1.54 Å, from the slipped intermediate geometry to a near-eclipsed reorganized geometry, H increases from 0.141 to 0.234 eV. This 66% enhancement shows that improved chain alignment strengthens interchain electronic coupling and facilitates electron hopping. Because the inter-chain electron transfer rate in Marcus theory scales as $H^2$, this geometric change alone translates into an approximately 2.75-fold acceleration of inter-chain hopping, independent of any change in the reorganization energy. The lateral slip is therefore not a passive structural detail but an active determinant of the inter-chain charge transport kinetics, and the spontaneous reorganization toward near-perfect π-overlap observed in the AIMD trajectory represents a doping-driven self-optimization of the transport network. This picture provides a coherent microscopic basis for the experimentally observed conductivity maximum in BBL near intermediate doping levels and underscores the intimate coupling between chain registry, polaron delocalization, and macroscopic electronic performance.

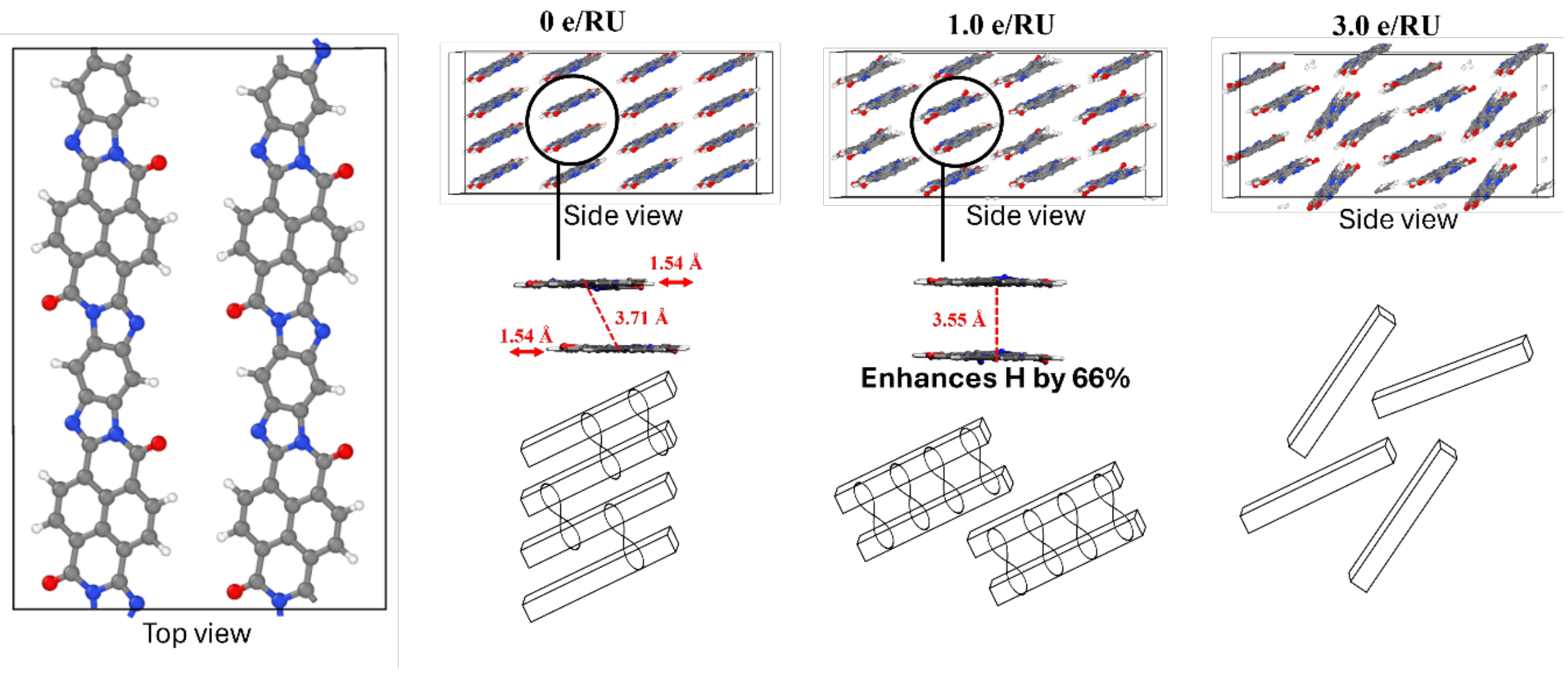

***Figure 8***. *Geometrical response of pristine BBL chains to electron doping. At low reduction levels, BBL chains form eclipsed shape with a significantly higher H value, a key Marcus parameter for rate of electron transfer.*

At the 2.0 e/RU reduction stage, the system unexpectedly recovers a high degree of order, with the chains aligning similarly to the neutral state, albeit with a minimal but non-negligible lateral offset. However, this stability is short-lived. At 2.5 e/RU, the system again loses order, forming pairs that exhibit a 'T-shape' overlap rather than the favorable π-stacking seen at lower densities. This structural integrity disappears almost entirely at 3.0 e/RU, requiring significantly more time for adjustment (4 ps vs. 2 ps). Matching this consecutive ordering-disordering cycle with the zigzag behavior of the bandgap suggests that while specific overlaps dominate transport at 1.0-1.5 e/RU, the resulting structural chaos at 3.0 e/RU renders the system chemically unsuitable for efficient electron transport. (Figure 8, right)

Using the AIMD trajectories, bond lengths were extracted for selected internal coordinates at each reduction level by averaging over the corresponding trajectory window after excluding the initial equilibration frames. To capture both the structural response and its dynamical variability, we computed the mean bond length and the RMSD for each bond within every charge state (Figure 9, top row). This analysis reveals a clear and progressive reorganization of the carbonyl-containing fused aza-ring framework upon electron addition.

The most direct signature of reduction is the monotonic elongation of the C=O bonds, whose average values increase from about 1.229 Å in the neutral state to about 1.270 Å at 3.0 e added charge, indicating substantial weakening of carbonyl double-bond character. This trend is accompanied by a redistribution of bonding within the adjacent fused rings. In particular, the C1–C2 bond shortens markedly from 1.512 to 1.450 Å, consistent with increased double-bond character, while the C1-N2 bond elongates from 1.330 to 1.367 Å, showing loss of imine-like character. At the same time, the C1-N3 and C4-N3 bonds lengthen more modestly, suggesting reduced N-mediated conjugation between the carbonyl carbon and the fused heterocyclic backbone. As a result, reduction is not localized solely on the carbonyl group, but instead drives a coordinated, delocalized rearrangement of bond order across the entire carbonyl–aza conjugated motif. The accompanying increase in bond-length fluctuations at higher charge further indicates that the reduced states are not only electronically distinct, but also more dynamically flexible.

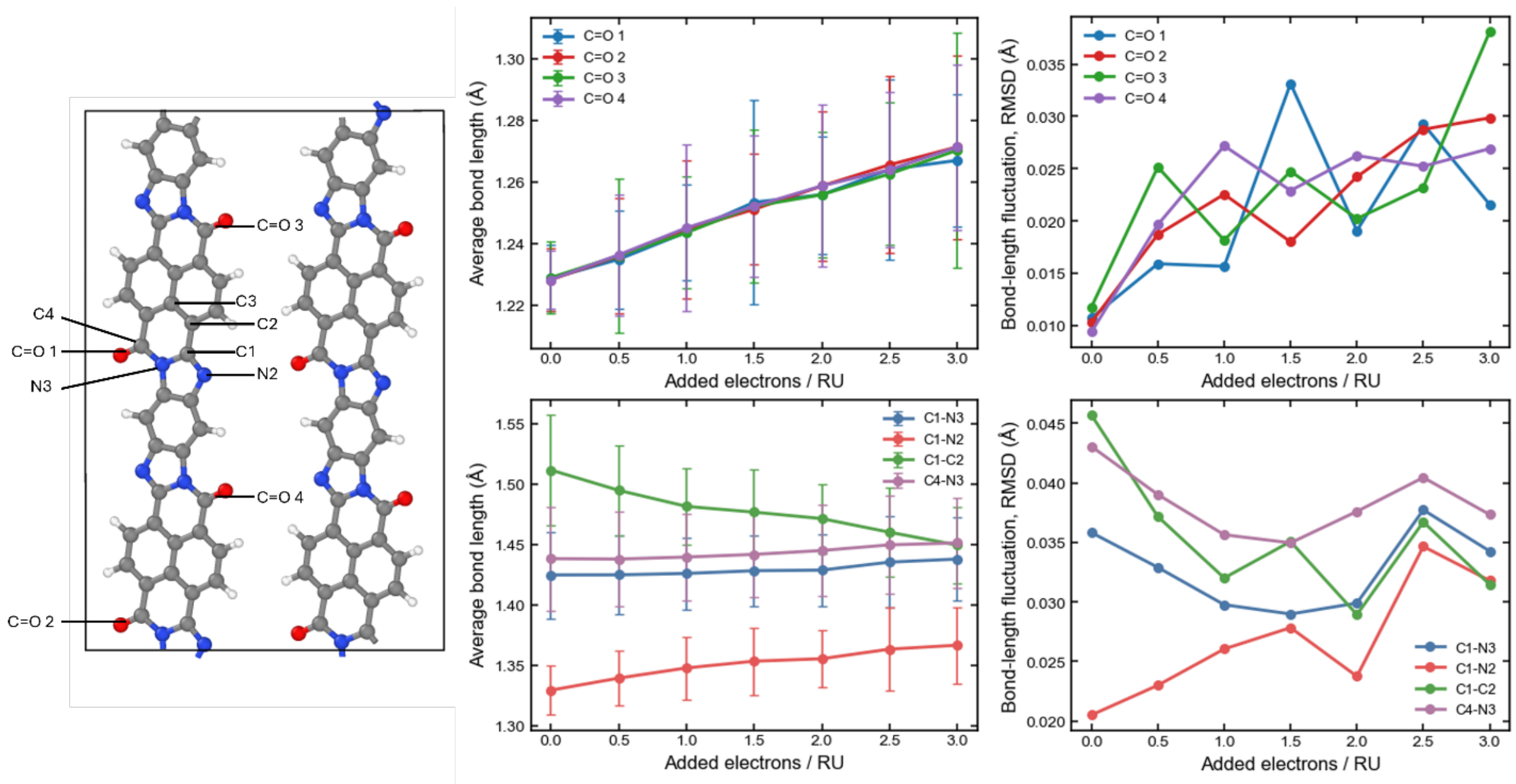

***Figure 9. Top:*** *Charge dependent evolution of the average C=O bond lengths (left) and their RMSD values (right).* ***Bottom:*** *Corresponding average bond lengths for selected backbone C–C and C–N bonds (left) and their RMSD values (right). The data show progressive weakening of carbonyl double-bond character upon reduction, accompanied by coordinated rearrangement of bonding within the fused conjugated backbone. The larger RMSD values at higher charge indicate that the reduced states are also more structurally dynamic.*

## 4. Ion effect

### 4.1. Effect of Explicit Electrolyte and Solvation on the Electronic Structure of BBL

Having established that the intrinsic electronic structure of BBL exhibits a strongly non-monotonic bandgap evolution with redox state, we next ask how this response is modified by the ionic environment present in electrochemically operated devices as a more realistic viewpoint. To address this, neutral KCl ion pairs per repetitive unit (KCl/RU) were systematically introduced into the simulations.  This leaves the polymer itself charge neutral and extracted bandgaps across all configurations in both the gas phase and within an implicit aqueous dielectric. This two-environment comparison is designed to isolate the competing contributions of direct ion–polymer electrostatic interactions and long-range dielectric screening.

The pristine BBL chain exhibits a bandgap of 1.79 eV in the gas phase, modestly reduced to 1.67 eV upon solvation, confirming that dielectric embedding alone produces only a minor perturbation of the frontier electronic structure. (Figure 10, top) The introduction of a single KCl pair, however, produces a substantially larger narrowing in both environments (1.30 eV in gas phase; 1.35 eV in water), indicating that even a single ion pair exerts a strong local electrostatic influence on the BBL π system. With increasing ion-pair loading, the two environments diverge sharply. In the gas phase, successive KCl addition leads to continued monotonic narrowing (1.26 eV at 2 KCl/RU; 1.03 eV at 3 KCl/RU), whereas in aqueous solution this trend reverses. At 2 KCl/RU the bandgap recovers to 1.67 eV, being comparable to the solvated pristine system, and remains relatively large at 1.46 eV for 3 KCl. (Figure 10, bottom)

This environment-dependent divergence has a clear microscopic origin. In the gas phase, the unscreened electrostatic fields of $K^+$ and $Cl^-$ progressively polarize the BBL π-electron system, destabilizing and reordering the frontier orbitals and narrowing the bandgap monotonically with increasing ion-pair concentration. In the aqueous environment, dielectric screening attenuates these long-range interactions; at higher electrolyte loadings, the occupied and unoccupied manifolds are stabilized more symmetrically, partially restoring the bandgap rather than continuing to narrow it. (Figure S2) The result is a non-monotonic, environment-gated response that closely mirrors the intrinsic zigzag behavior identified under direct reduction, suggesting that ionic doping and electrostatic screening engage the same frontier orbital reorganization mechanism, but with dielectric environment acting as an external control parameter.

Through this, the impact of ionic dopants on BBL is not determined by dopant concentration alone, but by the interplay between local ion–polymer interactions and environmental screening. This directly implicates device operation, where both parameters vary simultaneously. The charge density redistribution patterns underlying this response are examined in detail in the following section.

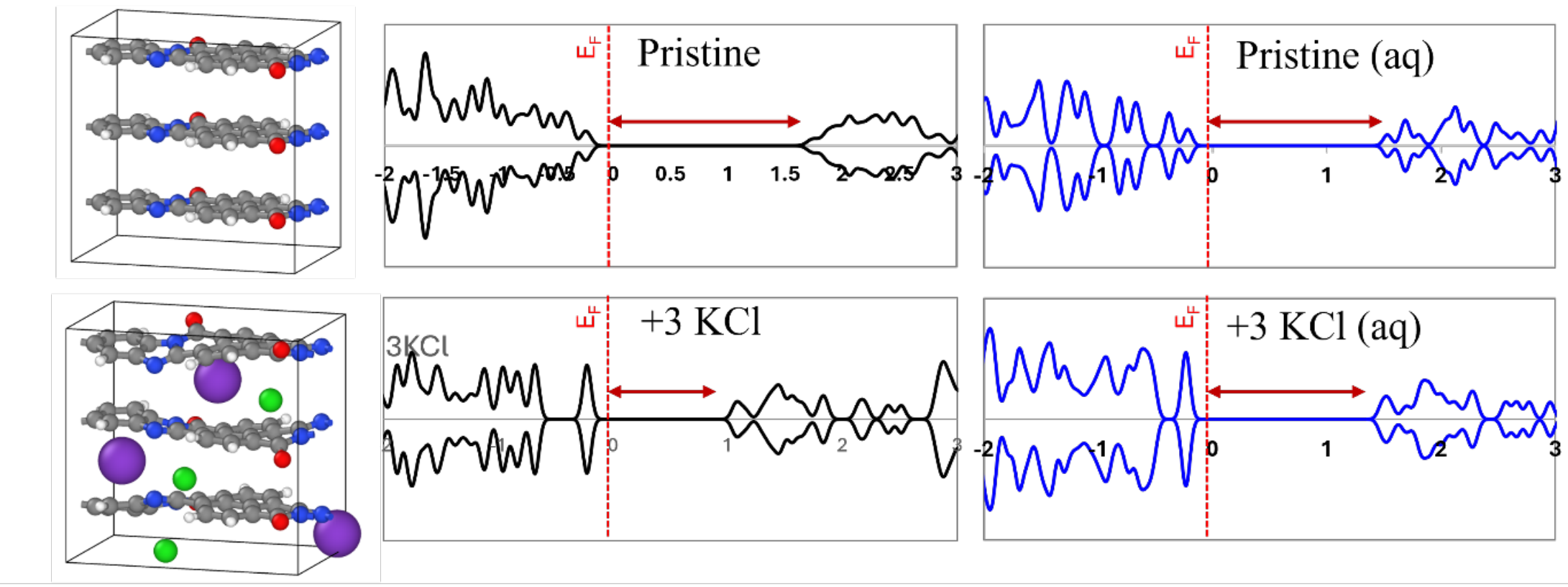

***Figure 10.*** *TDOS of pristine and 3 KCl-doped BBL in the gas phase and implicit water, with the Fermi level set to zero. While 3 KCl doping strongly narrows the bandgap in the gas phase, dielectric screening in water attenuates this frontier-orbital reorganization and partially restores the gap.*

#### 4.2. Charge Density Redistribution, Solvation Effects, and Bandgap Evolution

Charge density difference maps comparing implicit aqueous solvation to the gas phase show that ion-induced electronic redistribution in BBL is highly anisotropic and strongly dependent on dopant concentration. As a baseline, in the absence of electrolyte ions (0 KCl/RU) the solvent-gas density difference is distributed relatively evenly across all three vertically stacked BBL molecules, indicating a weak and delocalized solvation response. (Figure 11, left) Side-view projections along the polymer propagation direction reveal a subtle transverse asymmetry, with the charge density difference slightly rotating around the polymer direction. This behavior reflects an intrinsic anisotropic response of the BBL π system and provides a reference for assessing ion-induced effects.

For the 1KCl/RU system, where $K^+$ and $Cl^-$ reside between two of the three polymer layers, the solvent–gas density difference becomes strongly anisotropic and is predominantly localized on the side of the BBL backbone opposite to the ions. The redistribution is most pronounced at carbonyl (C=O) groups proximal to $K^+$, weaker at C=O groups near $Cl^-$, and minimal on the third, more distant layer aside from its carbonyls and minor contributions on select nitrogen atoms. Compared to the symmetric baseline at 0 KCl/RU, this demonstrates that ion pairing amplifies and localizes the intrinsic solvation-induced polarization.

At 2KCl/RU, the BBL backbone exhibits local curvature near the ions, with carbonyl groups adjacent to $Cl^-$ tilting out of plane, consistent with an electrostatically driven conformational response. Despite these distortions, the solvent-gas charge redistribution remains primarily on the side of the backbone opposite to the ions, indicating that dielectric screening continues to suppress ion-facing polarization. Carbonyl groups remain the dominant contributors to the solvent-dependent response.

The 3KCl/RU system displays more complex, non-additive behavior. While regions near $K^+$ continue to exhibit a twisted redistribution pattern that relocates electronic density to the opposite side of the backbone, polymer segments confined between closely spaced $Cl^-$ ions show strongly attenuated solvent-gas density differences, indicating that overlapping ionic fields and dielectric screening can locally suppress solvent-induced polarization. (Figure 11, right)

These redistribution patterns provide a microscopic context for the observed bandgap trends. In the gas phase, strong ion-facing polarization compresses frontier electronic levels, leading to monotonic bandgap narrowing with increasing KCl loading. In contrast, implicit solvation screens ionic electrostatic fields and suppresses this polarization, stabilizing a more balanced electronic distribution. At higher dopant concentrations, this suppression becomes dominant, resulting in a reopening of the bandgap in water.

Overall, the charge density difference analysis shows that electrolyte ions modulate the electronic structure of BBL through collective, environment-dependent polarization coupled to local structural reorganization. Solvation suppresses strong ion-facing distortions present in the gas phase, leading to redistribution of electronic density across the backbone and, at elevated dopant loadings, increased separation between occupied and unoccupied states. This solvent-mediated polarization provides a consistent microscopic explanation for the environment-dependent bandgap evolution observed upon electrolyte doping.

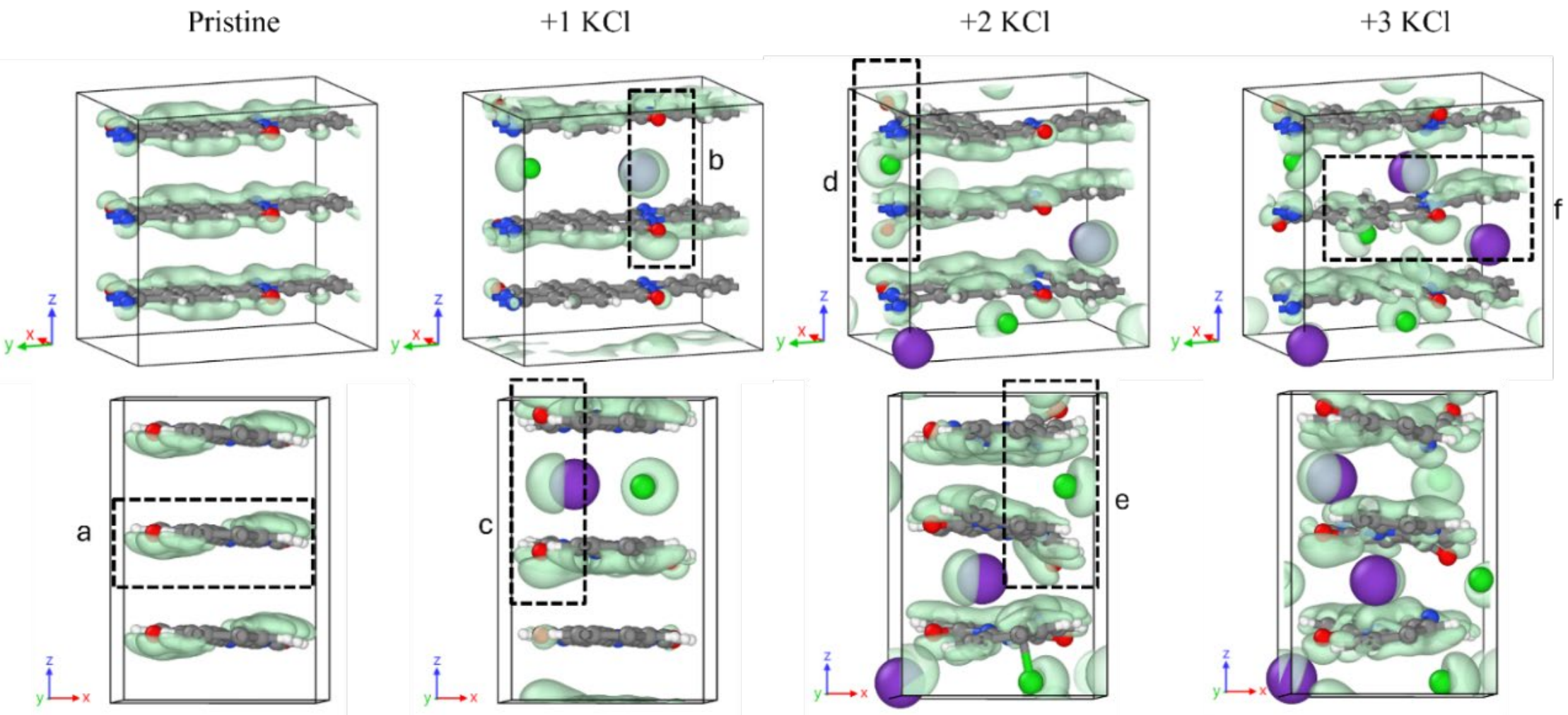

***Figure 11.*** *Charge density difference maps comparing implicit aqueous solvation with the gas phase for BBL with increasing KCl content, shown from left to right as pristine, 1KCl/RU, 2KCl/RU, and 3KCl/RU. The maps show that solvation-induced charge redistribution evolves from a weak, delocalized response in pristine BBL to increasingly anisotropic and ion-localized polarization with KCl doping, with the strongest changes occurring near carbonyl groups and ion-adjacent backbone regions.*

The structural evolution of the $K^+$-doped lattice is primarily defined by a progressive tilting mechanism that dictates the limits of charge transport. At low reduction levels (0.5-1.0 e/RU), the BBL chains maintain their initial orientation, adopting a stable "slip-stacked" geometry that facilitates the peak transport facility observed in this regime. As reduction proceeds, the lattice undergoes coordinated lattice dilation and trans-boundary migration at 1.5 e/RU, where chain pairs translate toward cell interfaces to accommodate the increasing ionic volume.

While these $K^+$ ions act as electrostatic anchors that prevent the T-shaped disorder observed in pristine BBL, (Figure 8) they eventually trigger a lattice failure at extreme doping. At 3.0 e/RU, the system undergoes a significant phase transition where the chains abandon their characteristic tilt angle, rotating to align parallel to the cell surface. This transition to a highly compressed, lamellar phase effectively decuples the π-systems of adjacent chains, providing a structural origin for the kinetic quenching observed at high doping levels. (Figure S7)

This structural dynamism is fundamentally mediated by the mobility of the counterion, as evidenced by our comparative analysis with KCl-doped BBL. In striking contrast to the $K^+$-only system, the KCl-doped lattice remains structurally static; it fails to exhibit the characteristic chain rotation or lamellar packing, instead maintaining an orientation similar to its neutral state even at 3.0 ERU. Bond statistics reveal that while $K^+$ ions in the KCl system continue to induce localized C=O bond stretching (confirmed by consistent elongation of carbonyls in close proximity of ~3.0-3.5 Å to the cations) the presence of $Cl^-$ suppresses the global structural response. (figure 12) We propose that the formation of $K^+$ ... $Cl^-$ ion pairs limits the effective potential of the cations, "pinning" them within the lattice and preventing the long-range electrostatic coordination required to drive the tilting observed in the unshielded $K^+$-doped system.

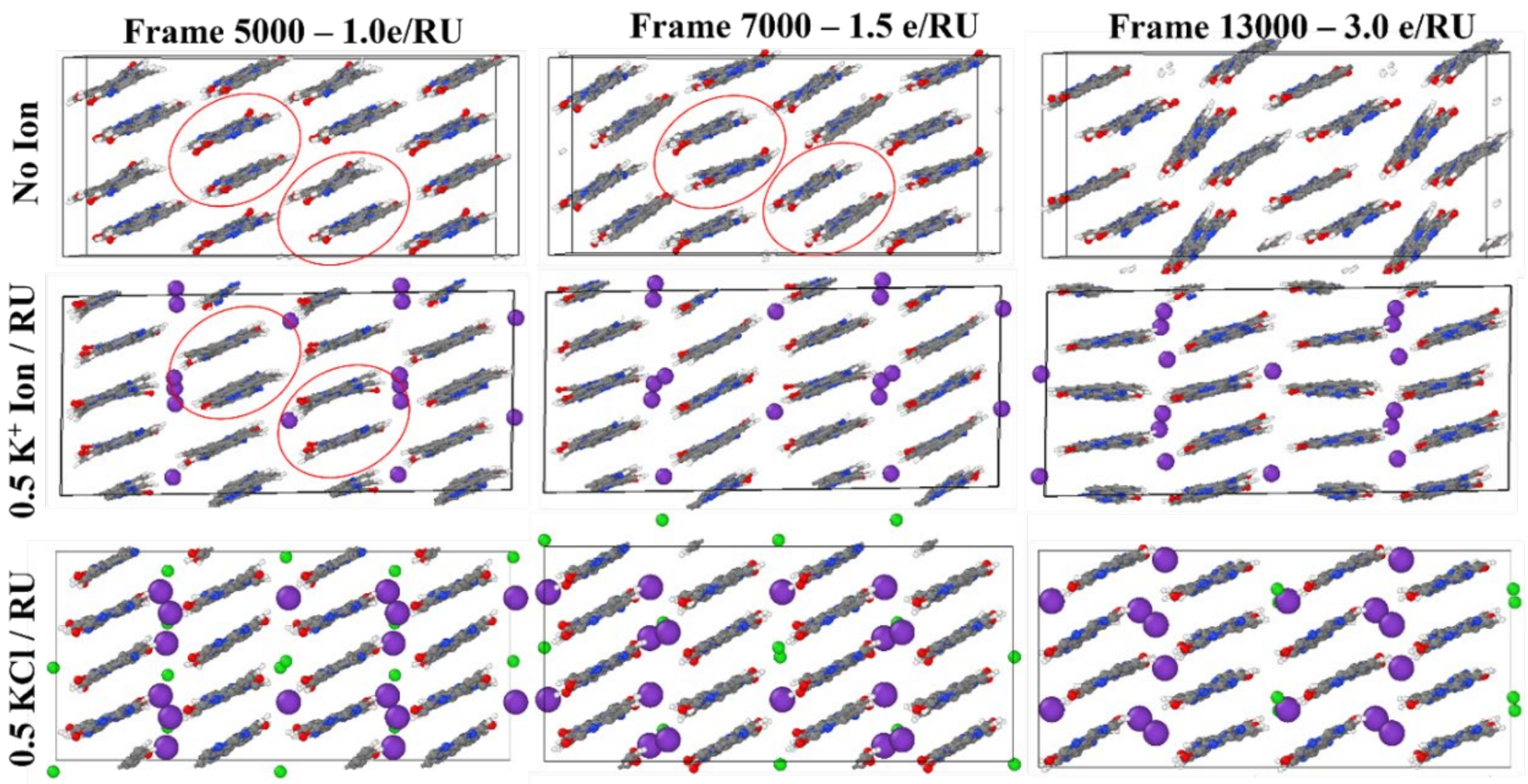

***Figure 12.*** *Side views of BBL chains at 1, 1.5, and 3 e/ru for the pristine system (top), with* $K^+$ *ions (middle), and with KCl salt (bottom). Ionic addition affects the structure beyond simple neutralization.*

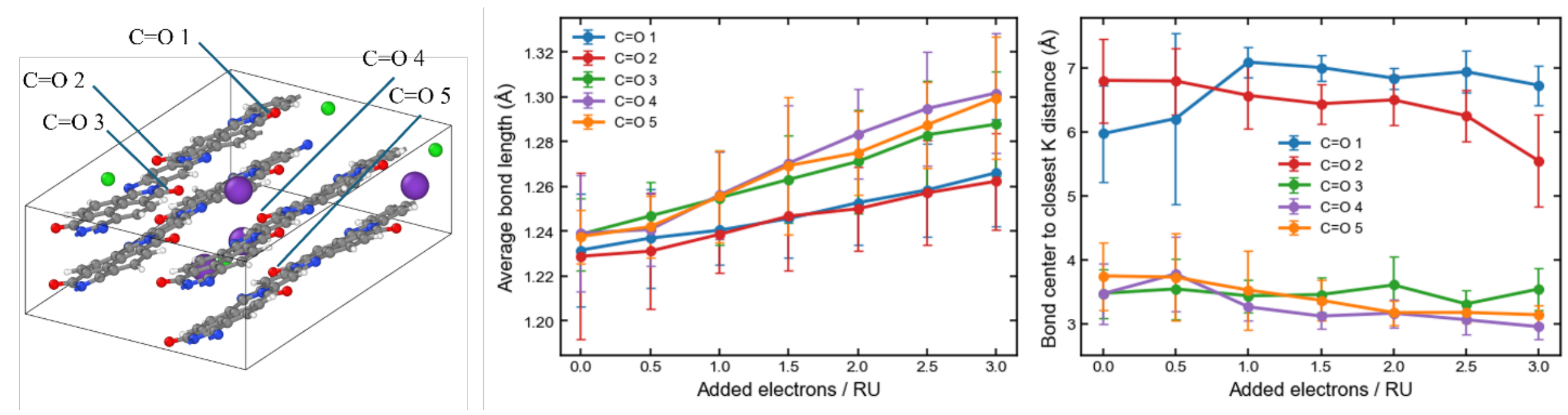

***Figure 13.*** *Average C=O bond lengths (left) and the corresponding* $K^+$ *proximity to the carbonyl groups (right). Their correlation indicates that closer* $K^+$ *coordination enhances electron localization on the C=O bonds during electron addition in AIMD.*

To probe the local electrostatic influence of the counterions, we tracked the geometric correlation between the $K^+$ positions and the BBL carbonyl C=O bond lengths. As shown in figure 13, top row, a direct linear relationship was discovered: C=O bonds in close proximity to a $K^+$ cation exhibit significant elongation, while symmetric $K^+$ coordination across multiple carbonyl groups results in uniform bond expansion. This site-specific polarization illustrates how counterions do not merely provide global charge neutrality but actively stabilize localized polaronic states through direct ion-dipole interactions. At the 3.0 e/RU limit, the saturation of these $K^+$ ... O=C interactions likely drives the observed backbone planarization and the transition to a sheared lamellar phase, as the lattice reorients to maximize cation-carbonyl coordination at the expense of long-range crystalline order.

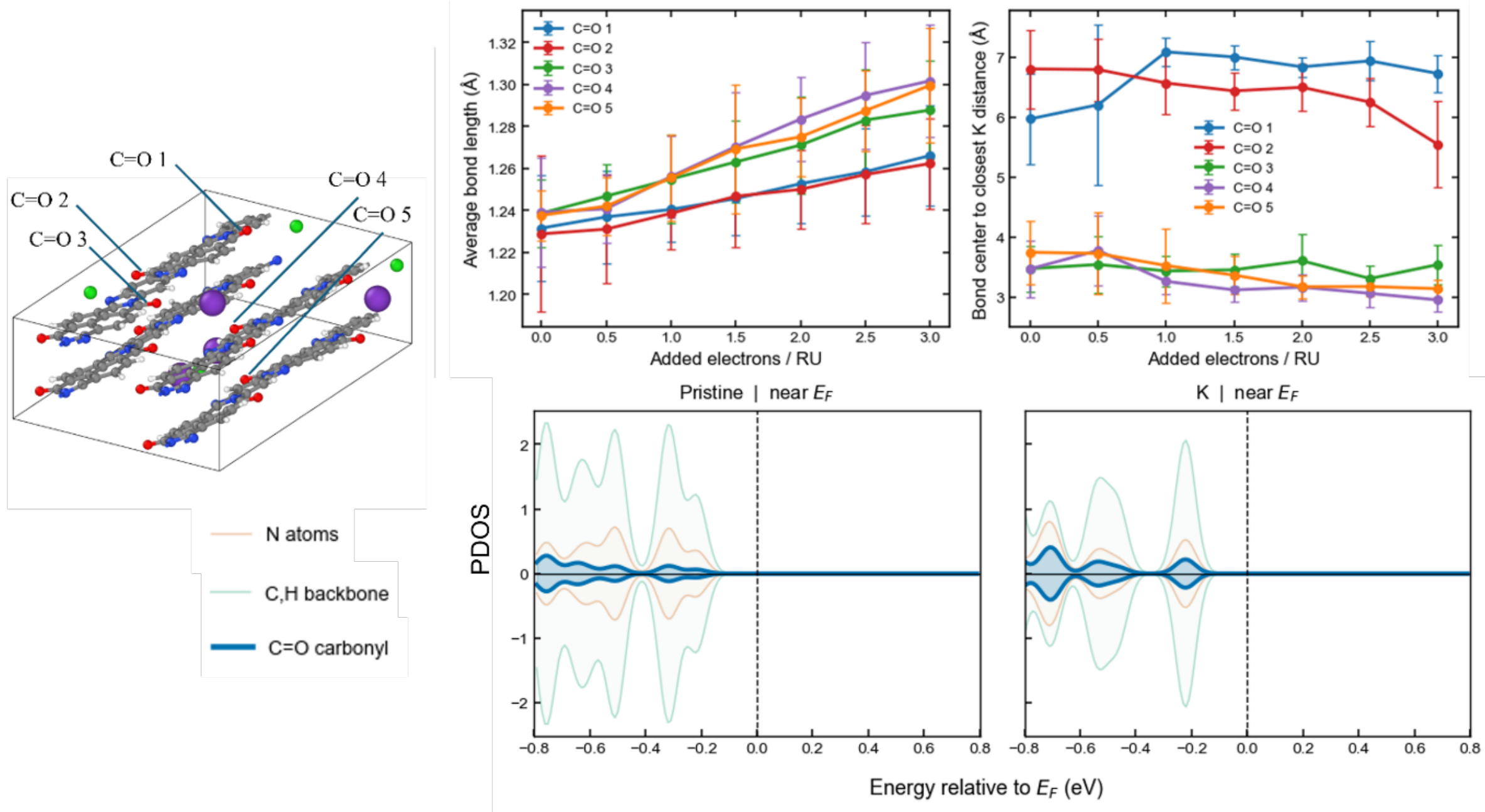

***Figure 14. Top:*** *Average C=O bond lengths (left) and the corresponding $K^+$ proximity to the carbonyl groups (right). Their correlation indicates that closer $K^+$ coordination enhances electron localization on the C=O bonds during electron addition in AIMD.* ***Bottom:*** *Zoomed-in PDOS of the pristine supercell compared with the KCl-containing system. $K^+$ cations promote greater charge localization on the carbonyl C=O groups. See* ***Figure S9*** *for full PDOS spectra.*

A comparison of the PDOS between the pristine and $K^+$ -containing cells reveals a fundamental reorganization of the frontier electronic structure. In the pristine system, the occupied frontier is dominated by backbone states, with carbonyl contributions remaining secondary (figure 13, bottom row). However, the introduction of $K^+$ ions induces a significant shift, bringing carbonyl character into the highest occupied bands before the Fermi level. This transition provides electronic evidence for **counterion-induced localization**, where the electrostatic field of the $K^+$ cation stabilizes carbonyl orbitals. This electronic 'activation' of the C=O sites correlate directly with our AIMD observations of localized bond stretching, suggesting that the counterions act as traps that redefine the charge-transport landscape through site-specific ion-dipole hybridization.

## 5. Conclusions

The antiambipolar behavior of poly(benzimidazobenzophenanthroline) (BBL) has been extensively characterized at the device level, yet whether this response reflects intrinsic material chemistry or emergent device physics has remained unresolved. The present work addresses this question through converging theoretical evidence that bell shaped conductivity in BBL originates in its fundamental electronic structure and supramolecular organization. Time dependent density functional theory (TDDFT) calculations reveal that the spatial delocalization of the first excited state evolves nonmonotonically across the doping series, progressing from low in the undoped system to markedly elevated at intermediate doping, before collapsing to highly localized states under heavy doping. This three regime progression maps directly onto the experimentally observed conductivity profile and establishes doping controlled modulation of charge transfer pathways as a defining feature of BBL's electronic landscape. Calculated Marcus rate of electron transfer ($k_{et}$) reinforces this picture with quantitative consistency: $k_{et}$ values are suppressed at both doping extremes and reach a well-defined maximum at intermediate concentrations, precisely reproducing the bell shaped functional form characteristic of antiambipolar transport. This rate maximum is further amplified by the solid state architecture of BBL, wherein molecular chains assemble into cofacial stacked layers that increase donor and acceptor frontier orbital overlap by 66%, substantially enhancing the electronic coupling that governs transfer efficiency. The cooperative alignment of optimal excited state localization and favorable intermolecular geometry at intermediate doping creates the conditions under which conductivity is maximized. The convergence of TDDFT analysis, Marcus theory, and orbital overlap characterization into a unified mechanistic picture compellingly demonstrates that the antiambipolar character of BBL is encoded in the intrinsic chemistry of the material, providing rigorous theoretical grounding for its continued use in antiambipolar device architectures.

## 6. Acknowledgements

This work was supported as part of the Reconfigurable Materials Inspired by Nonlinear Neuron Dynamics (reMIND) Energy Frontier Research Center, funded by the U.S. Department of Energy, Office of Science, Basic Energy Sciences at the Texas A&M Engineering Experiment Station under award #DE-SC0023353. Computational resources from the Texas A&M University High Performance Research Computing are gratefully acknowledged.

**Supporting Information for**

**Exploring the Origins of Anti-Ambipolarity in BBL Polymer: Links to Redox Chemistry, Electronic Structure, and Structural Dynamics**

**Maryam Ghotbi,[1] Alejandro Aviles,[2] and Perla B. Balbuena[1,2,3,*]**

**[1]Department of Chemistry, Texas A&M University, College Station, TX, USA**

**[2]Department of Chemical Engineering, Texas A&M University, College Station, TX, USA**

**[3]Department of Materials Science and Engineering, Texas A&M University, College Station, TX, USA**

*** e-mail: balbuena@tamu.edu**

A suitable functional for the BBL system: Studying the band gap of BBL is essential for establishing the energetic framework within which redox-induced charge carriers emerge and evolve. Variations in the band gap upon reduction reflect the formation of in-gap electronic states associated with polarons and bipolarons, which ultimately govern charge transport and conductivity. Accordingly, a comprehensive analysis of the band gap was performed across multiple redox states and environmental conditions. Prior to post-analysis, particular care was taken to ensure that the calculated electronic structure is consistent with experimental observations. Experimentally, the band gap of trans-BBL has been reported to be approximately 1.9 eV. To reproduce this value, we evaluated several combinations of exchange–correlation functionals and computational setups.

The PBE functional provides a computationally efficient description of structural trends in BBL; however, its semilocal nature introduces significant self-interaction error, which artificially stabilizes delocalized electronic states. As a consequence, PBE severely underestimates the band gap and, in extreme cases, predicts a vanishing gap, in disagreement with experiment (Figure 2a). This behavior has been reported previously for BBL and related conjugated polymers and highlights the inadequacy of semilocal functionals for describing their electronic structure. To address this limitation, dispersion-corrected approaches such as PBE-MBD have been proposed for extended π-conjugated systems. PBE-MBD augments PBE with a many-body dispersion correction, improving the description of long-range van der Waals interactions and structural organization. While the inclusion of many-body dispersion leads to the emergence of a finite band gap, the resulting value remains more than 1 eV smaller than the experimental reference, indicating that dispersion corrections alone are insufficient to remedy the electronic deficiencies of PBE.

As a third level of theory, we therefore employed the PBE0 hybrid functional to overcome the limitations of both PBE and PBE-MBD. While PBE and PBE-MBD offer reliable structural trends, both rely on semilocal exchange and consequently underestimate the band gap. In contrast, PBE0 incorporates a fixed fraction of exact exchange, which reduces self-interaction error and yields a more accurate separation between occupied and unoccupied electronic states. As a result, PBE0 reproduces the experimentally reported band gap of BBL with high fidelity. On this basis, PBE0 is adopted for subsequent electronic-structure analysis, providing a quantitatively reliable energetic reference for discussing charge transport in BBL.

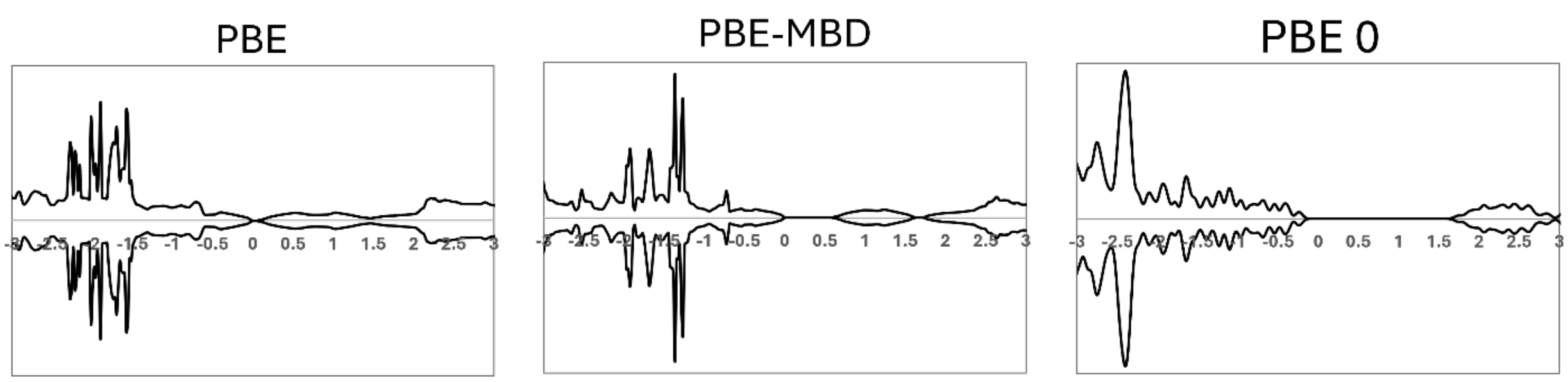

*Figure **S1** TDOS and extracted bandgaps of trans-BBL computed with PBE, PBE-MBD, and PBE0, compared with the experimental reference. While PBE predicts an unrealistically small gap and PBE-MBD only partially corrects it, PBE0 provides the closest agreement with experiment and a more reliable description of the frontier electronic structure.*

**Table S1** AIM Charges at the neutral ,and the singly doubly and triply reduced states for BBL Monomer.

| Element | Geom at Neutral St. | | | Neutral | 1 e/RU | 2 e/RU | 3 e/RU |
|---|---|---|---|---|---|---|---|
| C | 5.046158 | 5.106232 | 1.897418 | 0.956673 | 0.925564 | 0.839763 | 0.750648 |
| C | 5.308311 | 6.59319 | 1.892016 | -0.05883 | -0.07207 | -0.05434 | -0.06798 |
| C | 4.21603 | 7.520109 | 1.910699 | 0.016393 | -0.08845 | -0.05349 | -0.08117 |
| C | 2.901687 | 7.006898 | 1.934409 | -0.06192 | 0.006823 | -0.05267 | -0.05551 |
| C | 6.577547 | 7.155312 | 1.868082 | 0.008645 | -0.02758 | -0.10664 | -0.13704 |
| C | 4.421395 | 8.963949 | 1.907427 | -0.01123 | -0.05054 | -0.04998 | -0.07911 |
| C | 5.73582 | 9.477231 | 1.882604 | -0.05561 | -0.08467 | -0.10644 | -0.03831 |
| C | 6.773138 | 8.555977 | 1.861355 | 0.010733 | 0.031274 | -0.04044 | -0.11063 |
| C | 3.58991 | 11.37671 | 1.921871 | 1.0043 | 0.97498 | 0.81978 | 0.742678 |
| C | 3.328579 | 9.890109 | 1.926974 | -0.00897 | -0.01938 | -0.0603 | -0.02299 |
| C | 2.059654 | 9.327878 | 1.949959 | -0.01375 | -0.05013 | -0.00449 | -0.09106 |
| C | 1.864277 | 7.927985 | 1.954842 | 0.008388 | -0.02772 | -0.07688 | -0.13651 |
| C | 3.810384 | 3.172773 | 1.918432 | 0.288473 | 0.26982 | 0.196352 | 0.20586 |
| C | 5.208184 | 2.917079 | 1.896061 | 0.433398 | 0.412656 | 0.412309 | 0.407615 |
| C | 5.726366 | 1.589391 | 1.887473 | 0.103167 | 0.067374 | -0.0473 | -0.10476 |
| C | 4.825066 | 0.509756 | 1.901217 | 0.274834 | 0.257327 | 0.256546 | 0.267292 |
| C | 3.427258 | 0.765241 | 1.924424 | 0.390886 | 0.368942 | 0.41629 | 0.416456 |
| C | 2.909023 | 2.092876 | 1.93304 | 0.031092 | -0.00618 | -0.04217 | -0.10518 |
| C | 2.608701 | 5.507888 | 1.938825 | 1.39466 | 1.31569 | 1.20976 | 1.12475 |
| C | 6.026947 | 10.97687 | 1.877748 | 1.39093 | 1.3517 | 1.22949 | 1.09212 |
| H | 0.855761 | 7.540401 | 1.976979 | 0.076162 | 0.058159 | 0.020528 | 0.003288 |
| H | 1.199301 | 9.989685 | 1.964921 | 0.113447 | 0.083975 | 0.043862 | 0.039136 |
| H | 7.438368 | 6.493729 | 1.852786 | 0.111424 | 0.087541 | 0.067173 | 0.057094 |
| H | 7.782102 | 8.943511 | 1.837067 | 0.121209 | 0.048237 | 0.064812 | 0.051506 |
| H | 6.797308 | 1.437173 | 1.8708 | 0.097943 | 0.072259 | 0.075889 | 0.044405 |
| H | 1.83823 | 2.245933 | 1.950384 | 0.100585 | 0.075257 | 0.077983 | 0.04925 |
| N | 5.915853 | 4.130006 | 1.883404 | -1.09476 | -1.1489 | -1.17143 | -1.20793 |
| N | 3.717961 | 4.624412 | 1.919867 | -1.16579 | -1.17931 | -1.14278 | -1.15525 |
| N | 4.917662 | 11.85908 | 1.8993 | -1.14966 | -1.16973 | -1.1932 | -1.20851 |
| N | 2.720069 | 12.35236 | 1.936986 | -1.10987 | -1.16236 | -1.15216 | -1.21001 |
| O | 1.476414 | 5.081835 | 1.957967 | -1.09751 | -1.15589 | -1.19488 | -1.2232 |
| O | 7.159253 | 11.40373 | 1.856254 | -1.10544 | -1.16469 | -1.18095 | -1.21695 |

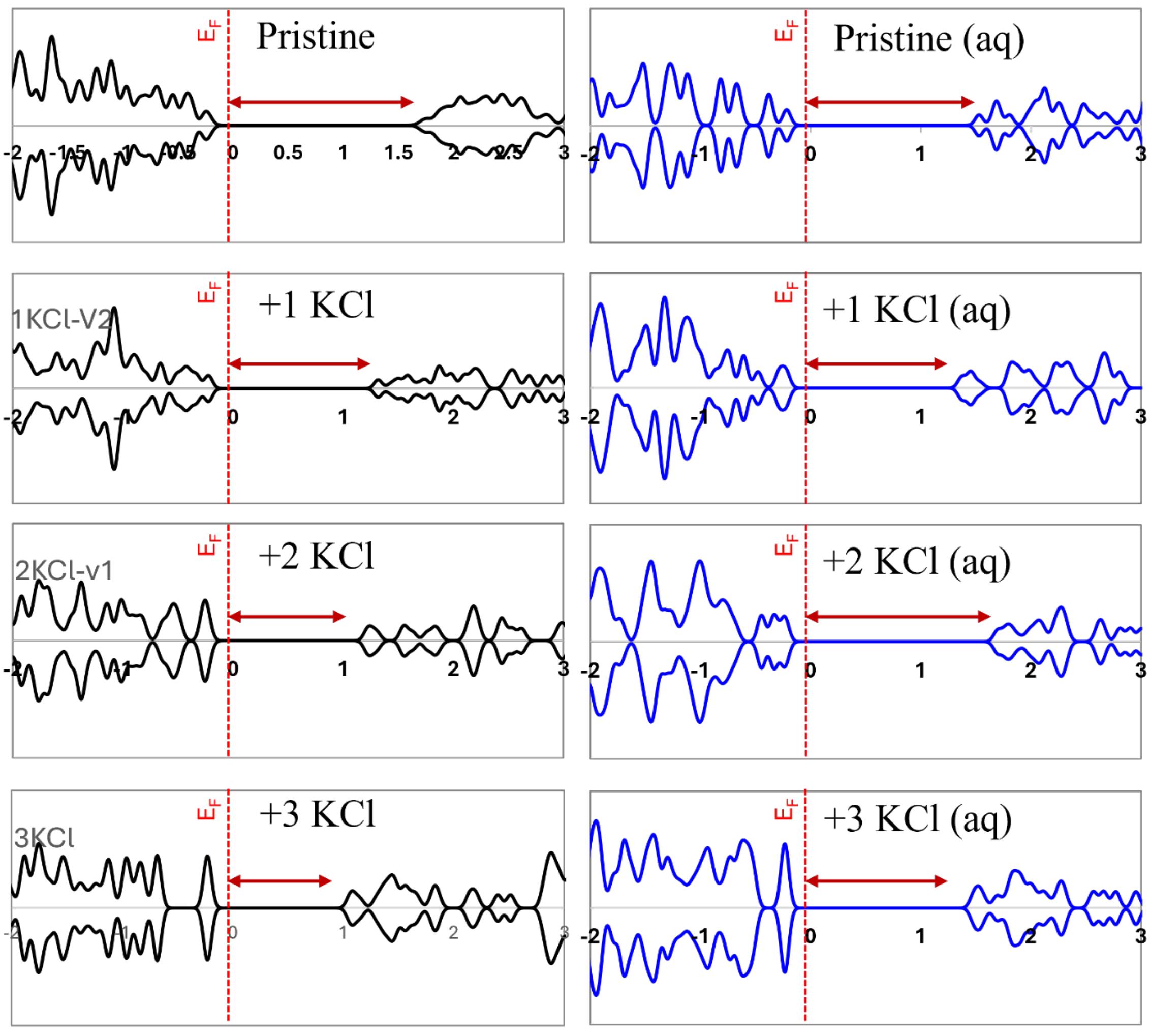

***Figure S2*** *Addition of counterions does more than provide charge balance. As the KCl concentration increases from 0 KCl/RU in the pristine cell (top row) to 1 KCl/RU (bottom row), the bandgap progressively decreases. This effect is less pronounced in implicit water, where dielectric screening weakens direct ion-polymer electrostatic interactions and partially preserves the frontier-orbital separation.*

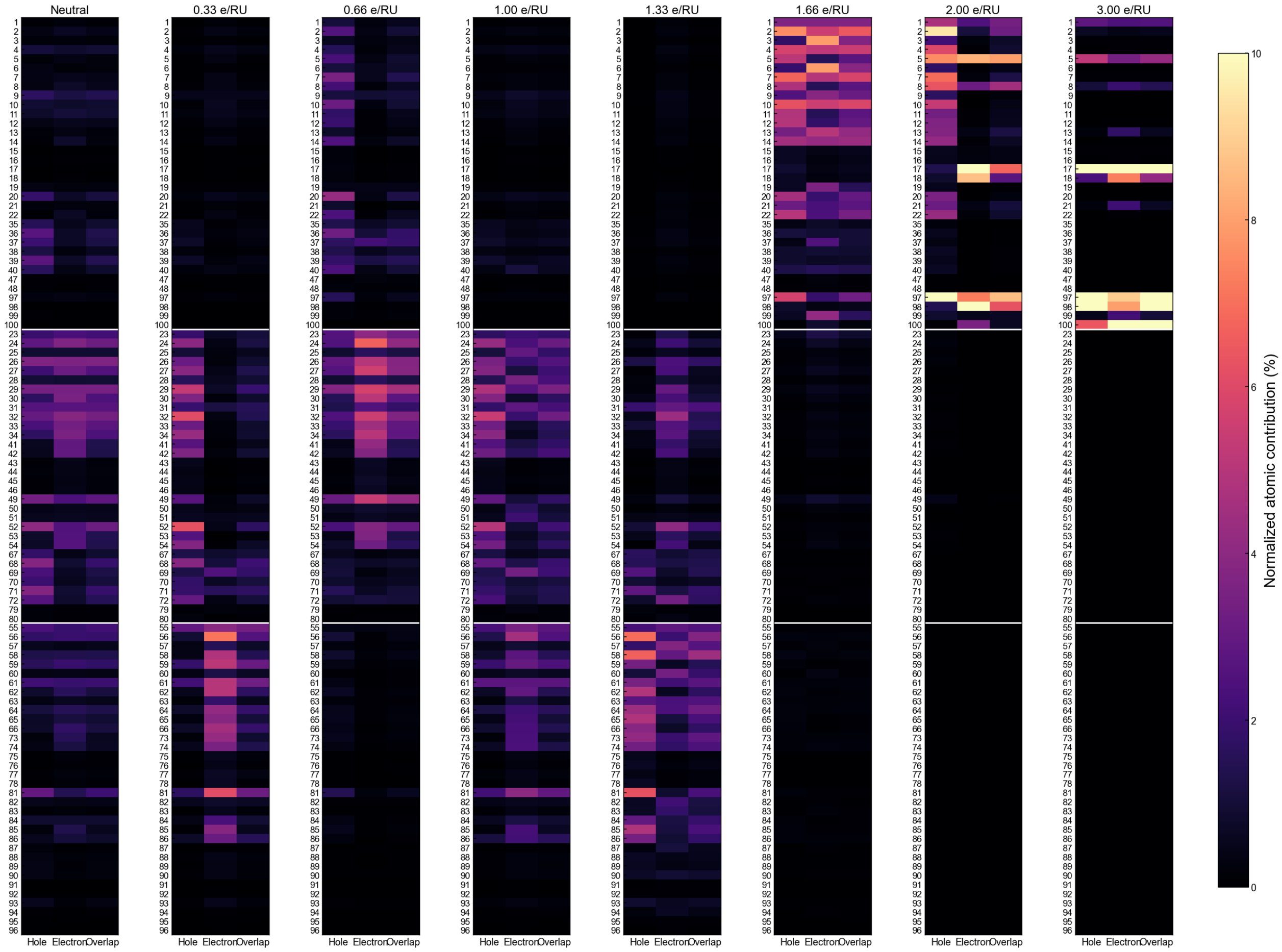

***Figure S3.*** *Atom-resolved donor and acceptor contributions for the first excited state of the BBL trimer with 0 to 3 e/RU added. The color scale represents the normalized atomic contribution to the hole and electron distributions, showing how donor and acceptor character is distributed across individual atoms.*

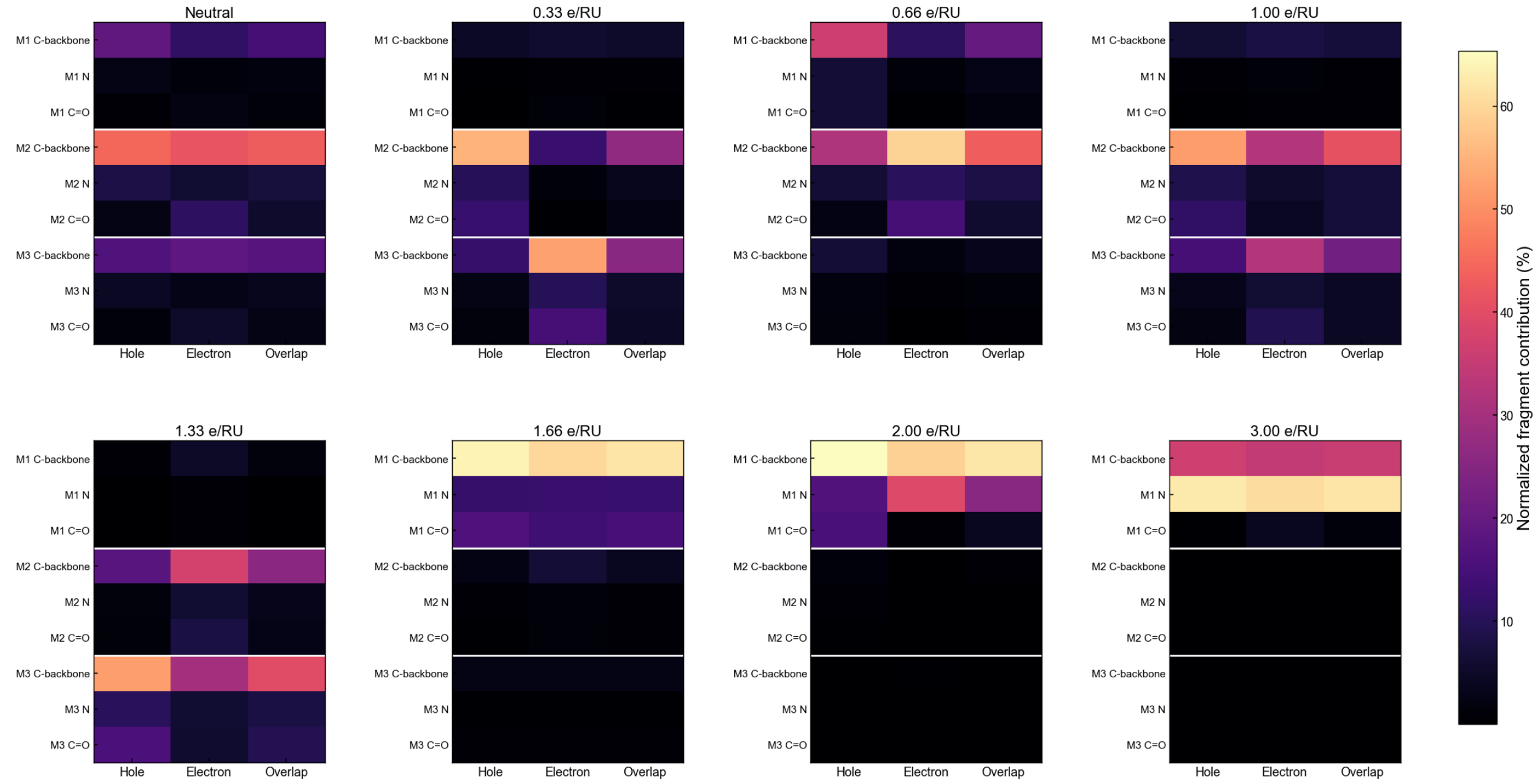

***Figure S4.*** *Fragment-resolved donor and acceptor contributions for the first excited state of the BBL trimer with 0–3 e/RU added. Contributions are grouped by monomer and chemical fragment, providing a clearer view of how hole and electron character is partitioned among the backbone carbon, nitrogen, and carbonyl-containing units of M1, M2, and M3, and how this partitioning evolves with increasing reduction.*

**Table S2.** Marcus parameters for all computed single-electron hops in the BBL trimer. Rows with missing $H$, $\lambda$, or $k_{et}$were omitted. The regime classification is based on the ratio $|\ \Delta G\ |/\lambda$: normal ( $|\ \Delta G\ |/\lambda < 0.9$ ), near-optimal $(0.9–1.1)$, and inverted ( $|\ \Delta G\ |/\lambda > 1.1$ ).

| Hop | ΔG (eV) | H (eV) | λ (eV) | ($k_{et}$ ($s^{-1}$) | Regime |
|---|---|---|---|---|---|
| e1 | -0.177 | 0.0019 | 0.817 | 5.09E+08 | normal |
| e2-1 | -3.580 | 0.3516 | 0.992 | 6.05E-14 | inverted |
| e2-2 | -0.627 | 0.1628 | 0.844 | 2.82E+14 | normal |
| e2-3 | -1.724 | 0.0164 | 1.610 | 3.28E+12 | near-optimal |
| e3-1 | -5.467 | 0.7992 | 0.602 | 7.95E-151 | inverted |
| e3-2 | -0.269 | 0.4063 | 0.772 | 1.30E+14 | normal |
| e3-3 | -1.494 | 0.0573 | 1.246 | 3.05E+13 | inverted |
| e3-4 | -1.716 | 0.7354 | 0.130 | 2.61E-66 | inverted |

| Hop | ΔG (eV) | H (eV) | λ (eV) | ($k_{et}$ ($s^{-1}$)) | Regime |
|---|---|---|---|---|---|
| e3-5 | -0.703 | 0.2686 | 0.056 | 8.18E-17 | inverted |
| e3-6 | -3.625 | 0.0018 | 0.851 | 3.57E-28 | inverted |
| e4-1 | -6.434 | 0.0439 | 0.960 | 4.81E-119 | inverted |
| e4-2 | -2.088 | 0.0122 | 0.755 | 3.26E+02 | inverted |
| e4-3 | -3.747 | 0.0167 | 1.242 | 1.85E-09 | inverted |
| e4-4 | -4.455 | 1.6520 | 1.050 | 9.99E-31 | inverted |
| e4-5 | -1.812 | 0.1093 | 0.992 | 2.78E+11 | inverted |
| e4-6 | -2.678 | 1.0492 | 1.063 | 7.70E+05 | inverted |
| e4-7 | -1.542 | 0.0003 | 0.814 | 2.89E+06 | inverted |
| e4-8 | -0.644 | 1.6508 | 0.061 | 3.34E-07 | inverted |
| e4-9 | -5.619 | 0.0000 | 0.557 | 9.20E-188 | inverted |
| e4-10 | -4.765 | 0.2191 | 1.494 | 3.39E-16 | inverted |
| e5-1 | -8.738 | 0.0137 | 0.898 | 1.28E-277 | inverted |
| e5-2 | -5.321 | 0.8427 | 0.943 | 1.53E-70 | inverted |
| e5-3 | -3.815 | 0.0001 | 0.739 | 2.21E-46 | inverted |
| e5-4 | -3.694 | 0.1724 | 2.306 | 9.69E+10 | inverted |
| e5-5 | -1.102 | 0.1017 | 0.787 | 5.75E+13 | inverted |
| e5-6 | -0.280 | 0.2841 | 0.812 | 5.07E+13 | normal |
| e6-1 | -10.016 | 0.5215 | 0.231 | 0.00E+00 | inverted |
| e6-2 | -7.469 | 1.1430 | 0.935 | 2.65E-177 | inverted |
| e6-3 | -2.673 | 0.8648 | 1.121 | 9.87E+06 | inverted |
| e6-4 | -6.315 | 0.0000 | 3.495 | 1.40E-04 | inverted |
| e6-5 | -2.426 | 0.0001 | 0.686 | 8.26E-11 | inverted |

| Hop | ΔG (eV) | H (eV) | λ (eV) | ($k_{et}$ ($s^{-1}$)) | Regime |
|---|---|---|---|---|---|
| e6-6 | -3.043 | 1.3326 | 0.647 | 1.06E-21 | inverted |
| e6-7 | -2.553 | 0.0757 | 0.948 | 3.19E+02 | inverted |
| e6-8 | -5.757 | 0.0000 | 0.727 | 1.23E-140 | inverted |

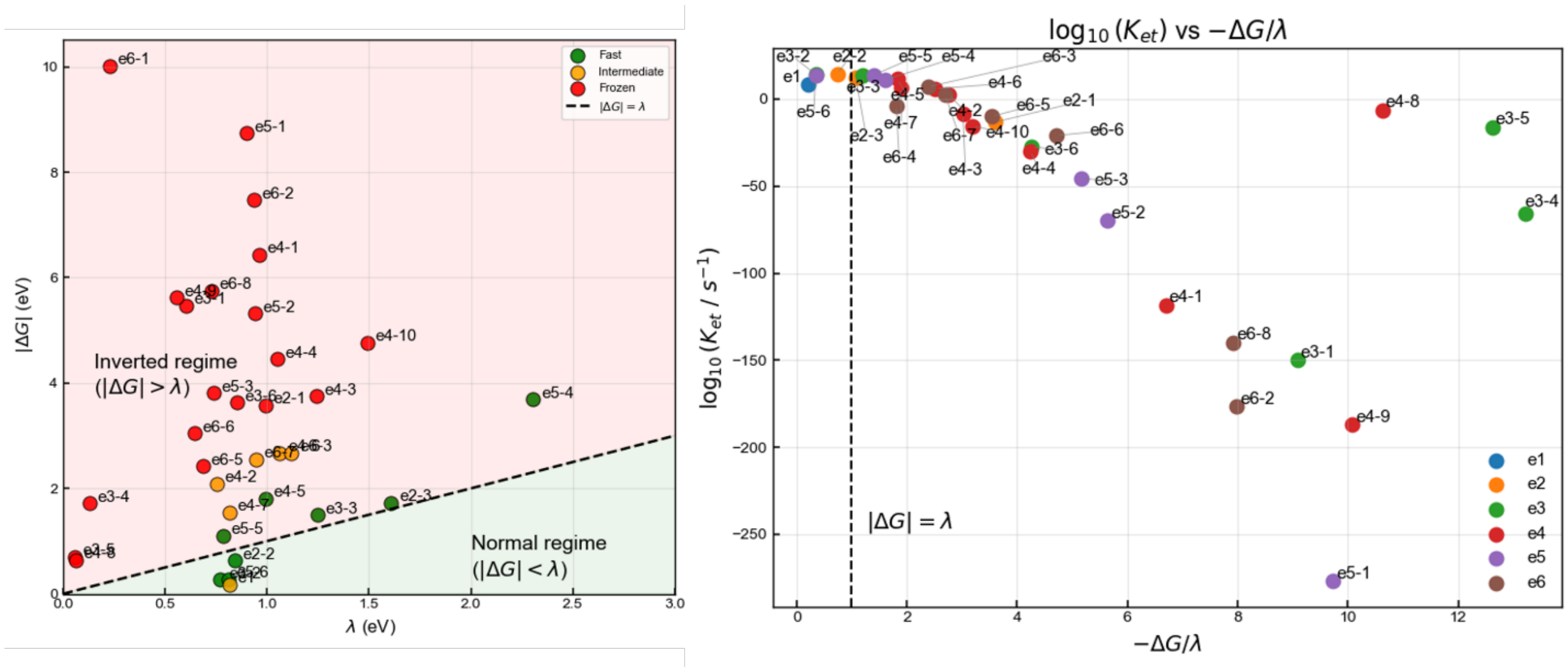

***Figure S5*** *Marcus parameters in harmony**. Left:** Marcus regime map for all computed single-electron hops in the BBL trimer across reduction states $n = 1$–$6$, plotted as $| \Delta G |$ versus $\lambda$. The dashed diagonal marks the Marcus optimum ($| \Delta G | = \lambda$). Fast hops cluster near this line, whereas highly exergonic pathways with $| \Delta G | \gg \lambda$ fall deep in the inverted regime and become kinetically suppressed.* ***Right:*** *Dependence of the calculated electron-transfer rate on the Marcus driving-force ratio, shown as $log_{10}(k_{et})$ versus $-\Delta G/\lambda$. The dashed vertical line at $-\frac{\Delta G}{\lambda} = 1$ marks the optimum condition, separating the normal and inverted regimes. Together, the two panels show that the fastest hopping pathways occur near the Marcus optimum, while over-reduction drives many hops into the inverted regime, providing a mechanistic basis for the conductivity collapse at high doping.*

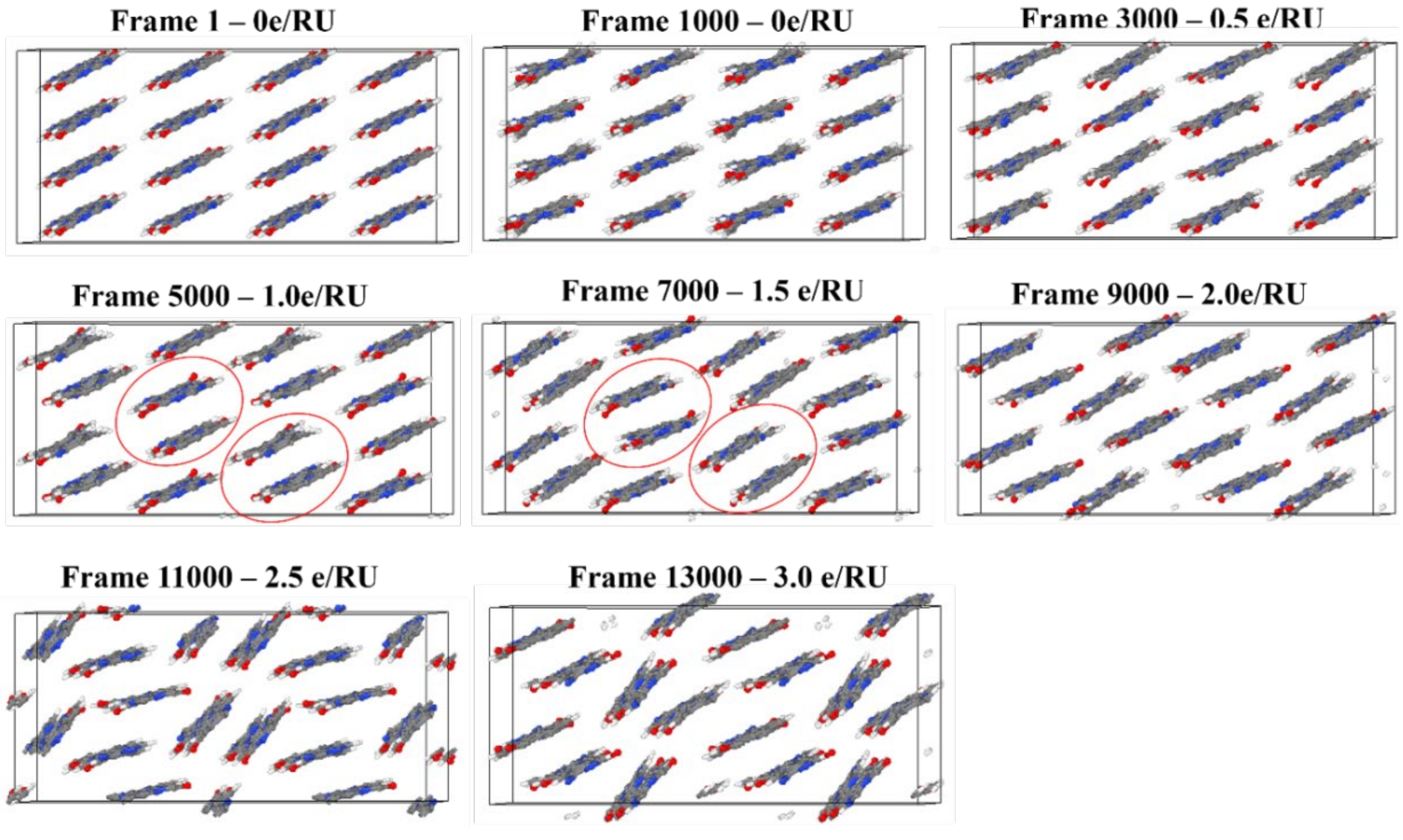

***Figure S6*** *Side views of the pristine BBL cell during the AIMD run with stepwise electron addition. Red circles indicate at 1 and 1.5 e/RU, the chains become highly stacked, whereas at 3 e/RU they lose their ordered arrangement.*

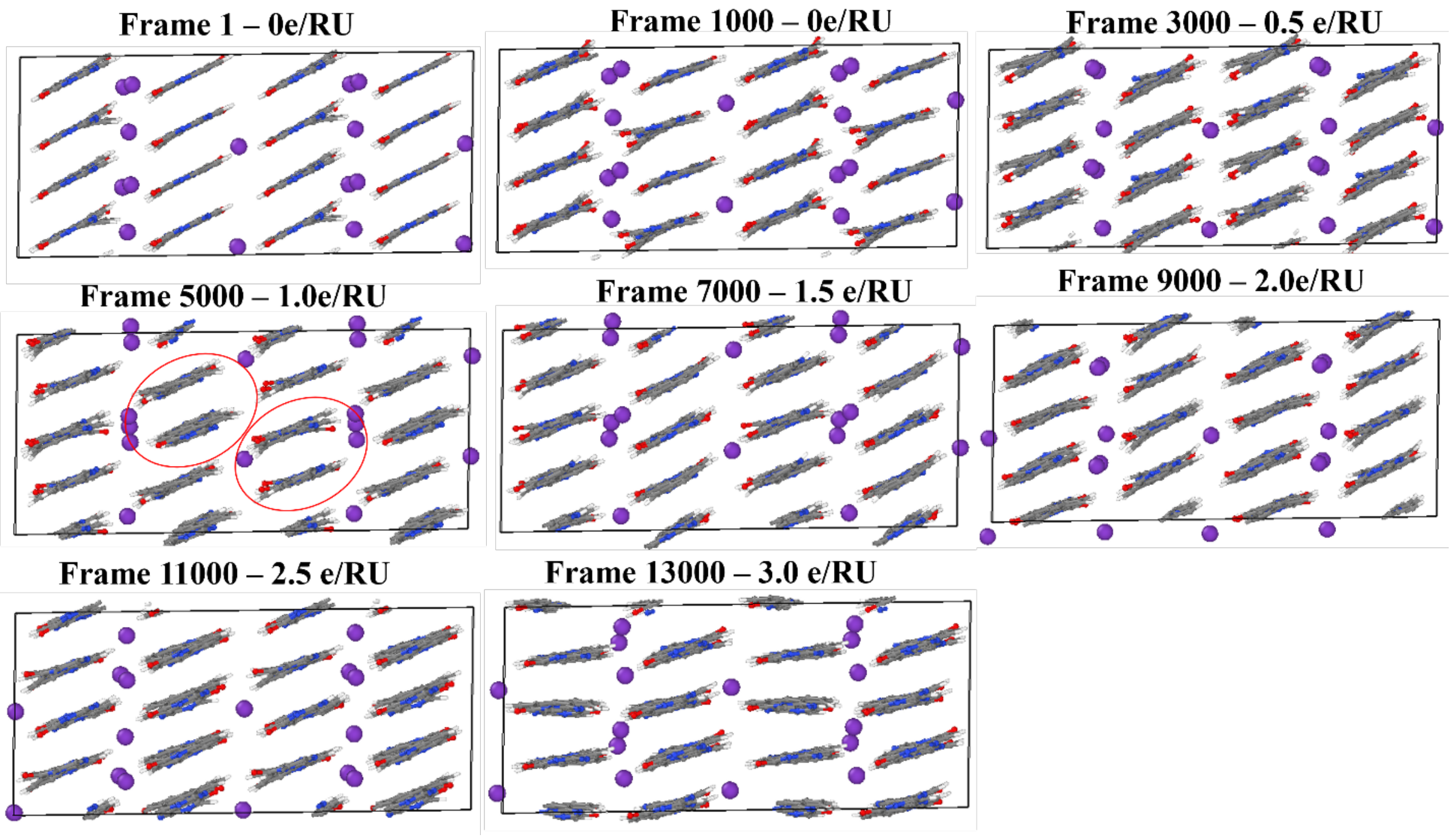

***Figure S7.*** *Representative AIMD snapshots of the $K^+$-doped BBL cell at selected times during stepwise reduction. Four electrons were added every 2000 ps, reaching 0-24 excess electrons. The sequence shows a transition from a stable slip-stacked geometry at low doping, favorable for charge transport, to lattice dilation and chain migration at intermediate reduction. $K^+$ ions suppress the T-shaped disorder seen in pristine BBL, but at 3.0 e/RU the lattice undergoes strong vertical shear and forms a lamellar phase, weakening interchain π-stacking and contributing to transport quenching.*

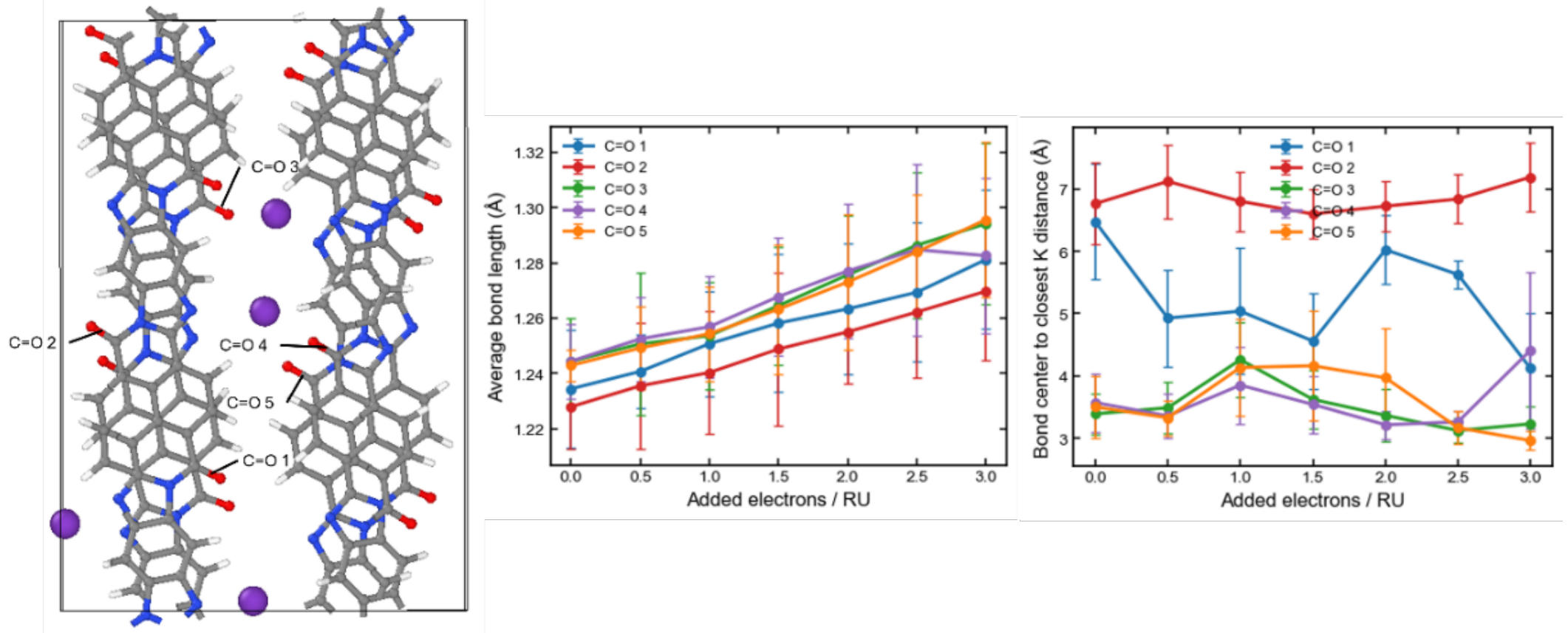

***Figure S8*** *Elongation of carbonyl double bonds as a response to doping. Closer $K^+$ ions promote electron addition to antibonding orbitals more significantly. Left: Average Bond length as a function of reduction State. Right: Distance of $K^+$ to carbonyl moiety.*

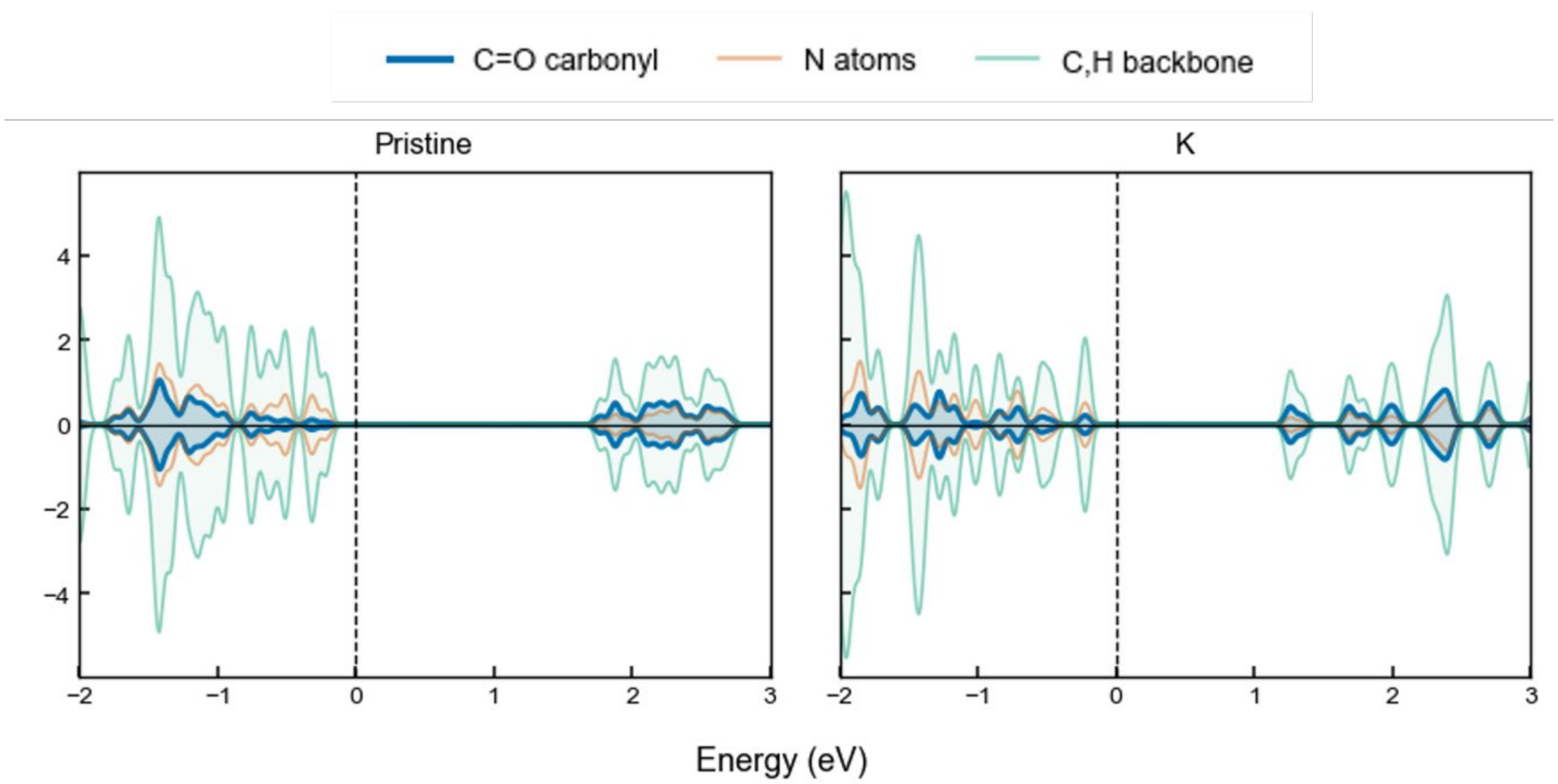

***Figure S9*** *Full PDOS spectra of Pristine BBL (left) and KCl-doped BBL (right). Addition of electrolyte promotes the activation of C=O groups as acceptors.*